\documentclass[aos,preprint]{imsart}

\RequirePackage[OT1]{fontenc}
\RequirePackage{amsthm,amsmath}
\RequirePackage{natbib}
\RequirePackage[colorlinks,citecolor=blue,urlcolor=blue]{hyperref}
\RequirePackage{graphicx}
\RequirePackage{amsthm}
\RequirePackage[cmex10]{amsmath}
\usepackage{amsmath}
\usepackage{amsfonts}
\usepackage{amssymb}
\startlocaldefs
\numberwithin{equation}{section}
\endlocaldefs
\theoremstyle{plain}
\newtheorem{theorem}{Theorem}[section]
\newtheorem{proposition}{Proposition}[section]
\newtheorem{corollary}{Corollary}[section]
\newtheorem{lemma}{Lemma}[section]
\newtheorem{assumption}{Assumption}
\newtheorem{definition}{Definition}
\theoremstyle{remark}
\newtheorem{remark}{Remark}[section]

\usepackage{amsfonts}
\usepackage{amsmath}	
\usepackage{bbm}	
\usepackage{easy-todo}	 

\newcommand{\ALR}{L}
\newcommand{\ee}{\varepsilon}
\newcommand{\ev}{v}
\newcommand{\score}{\phi}
\newcommand{\scoref}{\phi_f}

\newcommand{\Ze}{Z_{\varepsilon}}
\newcommand{\Zf}{Z_{\phi_f}}
\newcommand{\Zg}{Z_{\phi_g}}
\newcommand{\Zbk}{Z_{b_k}}
\newcommand{\Zgamma}{Z_{\Gamma}}

\newcommand{\We}{W_{\varepsilon}}
\newcommand{\Wf}{W_{\phi_f}}
\newcommand{\Wg}{W_{\phi_g}}
\newcommand{\Wbk}{W_{b_k}}
\newcommand{\Wb}{W_{b}}
\newcommand{\Wgamma}{W_{\Gamma}}

\newcommand{\WTe}{W^{(T)}_{\varepsilon}}
\newcommand{\WTf}{W^{(T)}_{\phi_f}}
\newcommand{\WTg}{W^{(T)}_{\phi_g}}
\newcommand{\WTga}{\widehat{W}^{(T)}_{\phi_g}}
\newcommand{\WTbk}{W^{(T)}_{b_k}}
\newcommand{\WTgamma}{W^{(T)}_{\Gamma}}
\newcommand{\Wind}{W_{\perp}}

\newcommand{\Be}{B_{\varepsilon}}
\newcommand{\Bf}{B_{\phi_f}}
\newcommand{\Bg}{B_{\phi_g}}
\newcommand{\BTg}{B^{(T)}_{\phi_g}}
\newcommand{\BTga}{\widehat{B}^{(T)}_{\phi_g}}
\newcommand{\Bbk}{B_{b_k}}
\newcommand{\Bb}{B_{b}}
\newcommand{\Bind}{B_{\perp}}

\newcommand{\EffCS}{\Delta^*}
\newcommand{\EffFI}{\mathcal{I}^*}
\newcommand{\EffCSg}{\Delta_g}
\newcommand{\EffFIg}{\mathcal{I}_g}
\newcommand{\EffCSgs}{\Delta_g^\mathbb{S}}
\newcommand{\EffFIgs}{\mathcal{I}_g^\mathbb{S}}
\newcommand{\EffCSind}{\Delta_\perp}
\newcommand{\EffCSinds}{\Delta_\perp^\mathbb{S}}
\newcommand{\EffCSe}{\Delta_\varepsilon}

\newcommand{\Ftf}{\phi} 
\newcommand{\limitE}{\mathcal{E}}
\newcommand{\limitEs}{\mathcal{E}_\mathbb{S}}
\newcommand{\limitET}{\mathcal{E}^{(T)}}
\newcommand{\limitETs}{\mathcal{E}^{(T)}_{\mathbb{S}}}
\newcommand{\maxInvf}{\mathcal{M}}	
\newcommand{\maxInvfs}{\mathcal{M}^{\mathbb{S}}}
\newcommand{\Inve}{\mathcal{M}_{\varepsilon}} 
\newcommand{\Invg}{\mathcal{M}_{g}}	
\newcommand{\Invgs}{\mathcal{M}_{g}^\mathbb{S}}
\newcommand{\coveg}{\sigma_{\varepsilon\phi_g}}

\newcommand{\statWTe}{\widehat{W}^{(T)}_\varepsilon}
\newcommand{\statWTind}{\widehat{W}^{(T)}_\perp}

\newcommand{\statBTg}{\widehat{B}^{(T)}_{\phi_g}}
\newcommand{\statBTind}{\widehat{B}^{(T)}_{\perp}}
\newcommand{\statLRmg}{\widehat{L}_{\Invg}}
\newcommand{\statLRmgs}{\widehat{L}_{\Invgs}}
\newcommand{\statLRg}{\widehat{L}_g}
\newcommand{\statLRgs}{\widehat{L}_g^\mathbb{S}}
\newcommand{\statCS}{\widehat{\Delta}_g}
\newcommand{\statCSs}{\widehat{\Delta}_g^\mathbb{S}}
\newcommand{\statFI}{\widehat{\mathcal{I}}_g}
\newcommand{\statFIs}{\widehat{\mathcal{I}}_g^\mathbb{S}}
\newcommand{\statDe}{\widehat{\Delta}_\varepsilon}
\newcommand{\statDind}{\widehat{\Delta}_\perp}
\newcommand{\statDinds}{\widehat{\Delta}_\perp^\mathbb{S}}

\newcommand{\statcoveg}{\hat\sigma_{\varepsilon\phi_g}}
\newcommand{\statJfg}{\hat{J}_{fg}}
\newcommand{\cvmg}{c_{\Invg}}
\newcommand{\cvmgs}{c_{\Invgs}}
\newcommand{\cvg}{c_{g}}
\newcommand{\cvgs}{c_{g}^\mathbb{S}}

\newcommand{\pto}{\stackrel{p}{\to}}
\newcommand{\wto}{\Rightarrow}
\newcommand{\law}[1]{\mathrm{P}^{(T)}_{#1}}
\newcommand{\prob}{\mathbb{P}}

\newcommand{\rE}[1][f]{\mathrm{E}_{#1}}
\newcommand{\indicator}[1]{\mathbbm{1}\left\{#1\right\}}
\newcommand{\var}{\operatorname{Var}}
\newcommand{\cov}{\operatorname{Cov}}
\newcommand{\rH}{\mathrm{H}}
\newcommand{\rd}{\mathrm{d}} 
\newcommand{\CS}{\Delta^{(T)}}
\newcommand{\CSlim}{\Delta}
\newcommand{\CSrho}{\Delta^{(T)}_{f}}
\newcommand{\CSlimrho}{\Delta_{f}}
\newcommand{\CSb}[1]{\Delta^{(T)}_{b_{#1}}}
\newcommand{\CSlimb}[1]{\Delta_{b_{#1}}}
\newcommand{\FIloc}[1][f]{J_{#1}} 
\newcommand{\FIfin}[1]{\mathcal{I}^{(T)}_{#1}}
\newcommand{\FIlim}{\mathcal{I}}
\newcommand{\FIaux}[1]{J_{#1}}
\newcommand{\SZ}{\mathbb{Z}} 
\newcommand{\SN}{\mathbb{N}} 
\newcommand{\SR}{\mathbb{R}} 
\newcommand{\F}{\mathfrak{F}} 
\newcommand{\Fs}{\mathfrak{F}_\mathbb{S}}
 
\newcommand{\Gspace}{\mathfrak{G}}
\newcommand{\rL}{\mathrm{L}}

\graphicspath {{figures/}}

\startlocaldefs
\endlocaldefs

\begin{document}

\begin{frontmatter}

\title{Semiparametrically Point-Optimal Hybrid Rank Tests for Unit Roots}
\runtitle{Semiparametrically optimal unit root tests}

%

\begin{aug}
\author{\fnms{Bo} \snm{Zhou}\thanksref{m1}\ead[label=e1,email]{bozhouhkust@ust.hk}},
\author{\fnms{Ramon} \snm{van den Akker}\thanksref{m2}\ead[label=e2,email]{r.vdnakker@gmail.com}}
\and
\author{\fnms{Bas J.M.} \snm{Werker}\thanksref{m2}
\ead[label=e3,email]{b.j.m.werker@tilburguniversity.edu}}
\runauthor{Zhou, van den Akker, and Werker.}
\affiliation{Hong Kong University of Science and Technology\thanksmark{m1} and Tilburg University\thanksmark{m2}}

%
%
%
\end{aug}

\begin{abstract}
We propose a new class of unit root tests that exploits invariance properties in the Locally Asymptotically Brownian Functional limit experiment associated to the unit root model. The invariance structures naturally suggest tests that are based on the ranks of the increments of the observations, their average, and an assumed reference density for the innovations. The tests are semiparametric in the sense that they are valid, i.e., have the correct (asymptotic) size,  irrespective of the true innovation density. For a correctly specified reference density, our test is point-optimal and nearly efficient. For arbitrary reference densities, we establish a Chernoff-Savage type result, i.e., our test performs as well as commonly used tests under Gaussian innovations but has improved power under other, e.g., fat-tailed or skewed, innovation distributions. To avoid nonparametric estimation, we propose a simplified version of our test that exhibits the same asymptotic properties, except for the Chernoff-Savage result that we are only able to demonstrate by means of simulations.
\end{abstract}


\end{frontmatter}

\section{Introduction}\label{sec:Introduction}
\noindent 
The monographs of \citep[][]{Patterson2011} (\citeyear{{Patterson2011}}, \citeyear{Patterson2012}) and \citet{Choi2015} provide an overview of the literature on unit roots tests. This literature traces back to \citet{White58} and includes seminal papers as \citeauthor{DickeyFuller79} (\citeyear{DickeyFuller79}, \citeyear{DickeyFuller81}), 
\citet{Phillips1987}, \citet{PhillipsPerron88}, and \citet{ERS1996}. The present paper fits into the stream of literature that focuses on ``optimal'' testing for unit roots. Important earlier contributions here are \citet{DufourKing91}, \citet{SaikkonenLuukkonen1993}, and \citet{ERS1996}. The latter paper derives the asymptotic power envelope for unit root testing in models with Gaussian innovations. \citet{RothenbergStock97} and \citet{Jansson2008} consider the non-Gaussian case.

The present paper considers testing for unit roots in a semiparametric setting. Following earlier literature, we focus on a simple AR(1) model driven by possibly serially correlated errors. The innovations driving these serially correlated errors are i.i.d., whose distribution is considered a nuisance parameter. Apart from some smoothness and the existence of relevant moments, no assumptions are imposed on this distribution. From earlier work it is known that the unit root model leads to Locally Asymptotically Brownian Functional (LABF) limit experiments \citep[in the Le Cam sense; see][]{Jeganathan1995}. As a consequence, no uniformly most powerful test exists (even in case the innovation distribution would be known) -- see also \citet{ERS1996}. In the semiparametric case the limit experiment becomes more difficult due to the infinite-dimensional nuisance parameter. \citet{Jansson2008} derives the semiparametric power envelope by mimicking ideas that hold for Locally Asymptotically Normal (LAN) models. However, the proposed test needs a nonparametric score function estimator which complicates its implementation. The point-optimal tests proposed in the present paper only require nonparametric estimation of a real-valued cross-information factor and we also provide a simplified version that avoids any nonparametric estimation. 

The main contribution of this manuscript is twofold. First, we derive the semiparametric power envelopes of unit root tests with serially correlated errors for two cases: symmetric or possibly non-symmetric innovation distributions (Section~\ref{sec:powerenvelope}). Our method of derivation is novel and exploits the invariance structures embedded in the semiparametric unit root model. To be precise, we use a ``structural'' description of the LABF limit experiment (Section~\ref{sec:limexp}), obtained from Girsanov's theorem. This limit experiment corresponds to observing an infinitely-dimensional Ornstein-Uhlenbeck process (on the time interval $[0,1]$). The unknown innovation density in the semiparametric unit root model takes the form of an unknown drift parameter in this limit experiment. Within the limit experiment, Section~\ref{sec:invariance} derives the maximal invariant, i.e., a reduction of the data which is invariant with respect to the nuisance parameters (that is, the unknown drift in the limiting Ornstein-Uhlenbeck experiment). It turns out that this maximal invariant takes a rather simple form: all processes associated to density perturbations have to be replaced by their associated bridges (i.e., consider $W(s) - sW(1)$ for the process $W(s)$ with $s\in[0, 1]$). The power envelopes for invariant tests in the limit experiment then readily follow from the Neyman-Pearson lemma. An application of the Asymptotic Representation Theorem (see, e.g., Theorem 15.1 in \citet{vdVaart00}) subsequently yields the local asymptotic power envelope (Theorem~\ref{thm:asymptotic_power_envelope}). In case the innovation density is known to be symmetric, the semiparametric power envelope coincides with the parametric power envelope. This implies the existence of an adaptive testing procedure (see also \citet{Jansson2008}). Moreover, we note that our analysis of invariance structures in the LABF experiment is also of independent interest and could, for example, be exploited in the analysis of optimal inference for cointegration or predictive regression models. Also, the analysis gives an alternative interpretation of the test proposed in \citet{ERS1996} --- the ERS test --- as this test is also based on an invariant, though not the maximal one (see Remark~\ref{remark:ERSlimit}).

As a second contribution, we provide two new classes of easy-to-implement unit root tests that are semiparametrically optimal in the sense that their asymptotic power curves are tangent to the associated semiparametric power envelopes (Section~\ref{sec:hybrid_rank_tests}). The form of the maximal invariant developed before suggests how to construct such tests based on the ranks/signed-ranks (depending on whether the innovation density is known to be symmetric or not) of the increments of the observations, the average of these increments, and an assumed reference density $g$. These tests are semiparametric in the sense that the reference density need not equal the true innovation density, while they are still valid (i.e., provide the correct asymptotic size). The reference density is not restricted to be Gaussian, which it generally needs to be in more classical QMLE results. When the reference density is correctly specified (i.e., $g$ happens to be equal to the true density $f$), the asymptotic power curve of our test is tangent to the semiparametric power envelope, and this in turn gives the optimality property. A feasible version of the oracle test using $g=f$ is obtained by using a nonparametrically estimated density $\hat{f}$, of which the corresponding simulation results are provided in Section~\ref{sec:MonteCarlo}. 

In relation to the classical literature on efficient rank-based testing (for instance, \citet{HajekSidak1967}, \citet{HallinPuri1988}, and \citet{HvdAW1}) our approach can be interpreted as follows. In the aforementioned papers, the invariance arguments (that is, using the ranks of the innovations) are applied in the sequence of models at hand. We, on the other hand, only apply the invariance arguments in the limit experiment. In this way, we can extend these ideas to non-LAN experiments. For the LAN case, both approaches would effectively lead to the same results. Our tests, despite the absence of a LAN structure, satisfy a \citet{ChernoffSavage1958} type result (Corollary~\ref{cor:chernoff-savage}): for any reference density our test outperforms, at any true density, its classical counterpart which, in this case, is the ERS test. We provide, in Section~\ref{sec:approximatehybridranktests}, even simpler alternative classes of tests that require no nonparametric estimations at all. These (simplified) classes of tests coincide with their corresponding originals for correctly specified reference density and, hence, share the same optimality properties. In case of misspecified reference density, the alternative classes still seem to enjoy the Chernoff-Savage type property, though only for a Gaussian reference density. This is in line with with the traditional Chernoff-Savage results for Locally Asymptotically Normal models.


The remainder of this paper is organized as follows. Section~\ref{sec:model} introduces the model assumptions and some notation. Next, Section~\ref{sec:powerenvelope} contains the analysis of the limit experiment. In particular we study invariance properties in the limit experiment leading to our new derivation of the semiparametric power envelopes. The classes of hybrid rank tests we propose are introduced in Section~\ref{sec:tests}. Section~\ref{sec:MonteCarlo} provides the results of a Monte Carlo study and Section~\ref{sec:conclusion} contains a discussion of possible extensions of our results. All proofs are organized in \ref{supplementA}.

\section{The model}\label{sec:model}
\noindent Consider observations $Y_1,\dots,Y_T$ generated from the classical component specification, for $t\in\SZ_{+}$,
\begin{align}
Y_t &= \mu + X_t, \label{eqn:DGP1} \\
X_t &=\rho X_{t-1}+\ev_t, \label{eqn:DGP2} \\
\Gamma(L)\ev_t &= \ee_t,  \label{eqn:DGP3}
\end{align}
where $\ev_0=\ev_{-1}=\cdots=\ev_{1-p}=0$, the innovations $\{\varepsilon_t\}$ form an i.i.d.\ sequence defined for $t\in\SZ$ with density $f$, and $\Gamma(L)$ is the AR(p) lag polynomial. Moreover, it is assumed that $Y_0 = \mu$. We impose the following assumptions on this innovation density.
\begin{assumption}\label{ass:density}\text{ }
\begin{enumerate}
\item[(a)] The density $f$ is absolutely continuous with a.e. derivative $f^\prime$, i.e., for all $a<b$ we have
\begin{align*}
f(b)-f(a)=\int_a^b f^\prime(e) \rd e.
\end{align*}
\item[(b)] $\rE[f] [\varepsilon_t]=\int e f(e) \rd e=0$ and
$\sigma_f^2=\var_f[\varepsilon_t]<\infty$. 
\item[(c)] The standardized Fisher-information for location,
\begin{align*}
\FIloc[f]=\sigma_f^2\int\scoref^2(e) f(e) \rd e,
\end{align*}
where $\scoref(e)= -(f^\prime /f)(e)$ is the  \textit{location score},
is finite.
\item[(d)]  The density $f$ is positive, i.e., $f>0$.  
\end{enumerate}
\end{assumption}
The imposed smoothness on $f$ is mild and standard (see, e.g., \citet{LeCam1986}, \citet{vdVaart00}). The finite variance assumption~(b) is important to our asymptotic results as it is essential to the weak convergence, to a Brownian motion,  of the partial-sum process generated by the innovations.\footnote{Let us already mention that, although not allowed for in our theoretical results, we will also assess the finite-sample performances of the proposed tests (Section~\ref{sec:MonteCarlo}) for innovation distributions with infinite variance. For tests specifically developed for such cases we refer to \citet{Hasan2001}, \citet{AhnFotopoulosHe03}, and \citet{CallegariCappuccioLubian03}.} The zero intercept assumption in (b) excludes a deterministic trend in the model. Such a trend leads to an entirely different asymptotic analysis, see \citet{HvdAW1}. The Fisher information $\FIloc[f]$ in~(c) has been standardized by premultiplying with the variance $\sigma_f^2$, so that it becomes scale invariant (i.e., invariant with respect to $\sigma_f$). In other words, $\FIloc[f]$ only depends on the shape of the density $f$ and not on its variance $\sigma_f^2$. The positivity of the density $f$ in (d) is mainly made for notational convenience.

The assumption on the initial condition, $\ev_0=\ev_{-1}=\cdots=\ev_{1-p}=0$, is less innocent then it may appear. Indeed, it is known, see \citet{MullerElliott2003} and \citet{ElliottMuller2006}, that, even asymptotically, the initial condition can contain non-negligible statistical information. Nevertheless, it is still stronger than necessary for the sake of simplicity, and it can be relaxed to the level of generality in \citet{ERS1996}.

Let $\F$ denote the set of densities satisfying Assumption~\ref{ass:density}. We also investigate in the present paper the special case of symmetric densities. For that purpose, we denote by $\Fs$ the set of densities which satisfy Assumption~\ref{ass:density} and, at the same time, are symmetric about zero. Of course, it follows $\Fs\subset\F$. 

With respect to the autocorrelation structure $\Gamma(L)$, we impose the following assumption.
\begin{assumption} \label{ass:gamma} 
	The lag polynomial $\Gamma(z) := 1-\Gamma_1 z - \cdots - \Gamma_p z^p$ is of finite order $p\in\SZ_{+}$ and satisfies $\min_{|z|\in\mathbb{C}:|z|\leq 1}|\Gamma(z)|>0$. 
\end{assumption}
Let $\Gspace\subset\SR^p$ denote the set of $\left(\Gamma_1,\ldots,\Gamma_p\right)'$ such that the induced lag polynomial satisfies Assumption~\ref{ass:gamma}. For mathematical convenience we restrict the lag polynomial to be of finite order $p$. One may expect many of the results in the present paper to extend to the case $p=\infty$ (see, e.g., \citet{Jeganathan1997}).

The main goal of this paper is to develop tests, with optimality features, for the semiparametric unit root hypothesis
\begin{align*}
\rH_0: \rho=1,\, (\mu\in\SR,\Gamma\in\Gspace,f\in\F) \text{~~vs~~}\rH_a:\, \rho<1,\, (\mu\in\SR, \Gamma\in\Gspace, f\in\F),
\end{align*}
i.e., apart from Assumption~\ref{ass:density}-\ref{ass:gamma}, no further structure is imposed on $f$, the intercept $\mu$, and the autocorrelation structure $\Gamma(L)$.

In the following section, we derive the (asymptotic) power envelope of tests that are (locally and asymptotically) invariant with respect to the nuisance parameters $\mu$, $\Gamma$, and $f$. We consider both the non-symmetric ($f\in\F$) and the symmetric case ($f\in\Fs$). Section~\ref{sec:tests} is subsequently devoted to tests, depending on a reference density $g$ that can be freely chosen, that are point optimal with respect to this power envelope and proves a Chernoff-Savage type result.

\section{The power envelope for invariant tests}\label{sec:powerenvelope}
\noindent This section first introduces some notations and preliminaries (Section~\ref{sec:prelim}). Afterwards, we will derive the limit experiment (in the Le Cam sense) corresponding to the component unit root model (\ref{eqn:DGP1})-(\ref{eqn:DGP3}) and provide a ``structural'' representation of this limit experiment (Section~\ref{sec:limexp}). In Section~\ref{sec:invariance} we discuss, exploiting this structural representation, a natural invariance restriction, to be imposed on tests for the unit root hypothesis with respect to the infinite-dimensional nuisance parameter associated to the innovation density. We derive the maximal invariant and obtain from this the power envelope for invariant tests in the limit experiment. At last, in Section \ref{sec:ART}, we exploit the Asymptotic Representation Theorem to translate these results to obtain (asymptotically) optimal invariant test in the sequence of unit root models. Again we consider both the case of unrestricted densities $f\in\F$ and that of symmetric densities $f\in\Fs$. 

\subsection{Preliminaries}\label{sec:prelim}
We first introduce local reparameterizations for the parameter of interest $\rho$ and the nuisance autocorrelation structure $\Gamma(L)$. Then we discuss a convenient parametrization of perturbations to the innovation density $f$ which we use to deal with the semiparametric nature of the testing problem. These perturbations follow the standard approach of local alternatives in (semiparametric) models commonly used in experiments that are Locally Asymptotically Normal (LAN). We will see that, with respect to all parameters but $\rho$, the model is actually LAN; compare also Remark~\ref{remark:perturbations} below. Moreover, we introduce some partial sum processes that we need in the sequel, as well as their Brownian limits. 

\subsubsection*{Local reparameterizations of $\rho$ and $\Gamma(L)$}
It is well-known, and goes back to \citet{Phillips1987}, \citet{ChanWei88} and \citet{PhillipsPerron88},
that the contiguity rate for the unit root testing problem, i.e., the fastest convergence rate at which it is possible to distinguish (with non-trivial power) the unit root $\rho=1$ from a stationary alternative $\rho<1$, is given by 
$T^{-1}$.  Therefore, in order to compare performances of tests with this proper rate of convergence, we reparametrize the autoregression parameter $\rho$ into its local-to-unity form, i.e.,
\begin{equation}\label{eqn:local_alternative}
\rho=\rho^{(T)}_h=1+\frac{h}{T}.
\end{equation}

The appropriate local reparameterization for the lag polynomial $\Gamma(L)$ is of a traditional form with rate $\sqrt{T}$, i.e.,

\begin{align} \label{eqn:local_gamma}
\Gamma^{(T)}_\gamma(L) = \Gamma(L) + \frac{\gamma(L)}{\sqrt{T}},
\end{align}
where the local perturbation $\gamma$ is defined by $\gamma(L) := - \gamma_1 L - \cdots - \gamma_p L^p$ with local parameter $\gamma := (\gamma_1,\dots,\gamma_p)'\in\SR^p$. As $\Gspace$ is open, $\Gamma^{(T)}_\gamma \in \Gspace$ for $T$ large enough. 

\subsubsection*{Perturbations to the innovation density}
\noindent
To describe the local perturbations to the density $f$, we  need the separable Hilbert space
\begin{align*}
\rL_2^{0,f}=\rL_2^{0,f}(\SR,\mathcal{B})
=\left\{  b\in \rL_2^f(\SR,\mathcal{B})\, \left|\,  
\int b(e) f(e) \rd e=0,\, \int e b(e)  f(e) \rd e
  =0\right.\right\}, 
\end{align*}
where $\rL_2^f(\SR,\mathcal{B})$ denotes the space of Lebesgue-measurable functions $b:\,\SR\to\SR$ satisfying $\int  b^2(e) f(e)\rd e<\infty$. Because of the separability, there exists a countable orthonormal basis $b_k$, $k\in\SN$, of $\rL_2^{0,f}$ (see, e.g., \citet[Theorem 3.14]{Rudin1987}). This basis can be chosen such that $b_k \in C_{2,b}(\SR) $, for all $k$, i.e., each $b_k$ is bounded and two times continuously differentiable with bounded derivatives. Moreover, $\rE b_k(\varepsilon)=0$ and $\var_f{b_k(\varepsilon)}=1$. Hence each function $b\in \rL_2^{0,f}$ can be written as $b= \sum_{k=1}^\infty  \eta_k b_k $, for some $\eta := (\eta_k)_{k\in\SN} \in \ell_2=\{ (x_k)_{k\in\SN}\, | \, \sum_{k=1}^\infty x_k^ 2<\infty\}$. Besides the sequence space $\ell_2$ we also need the sequence space $c_{00}$ which is defined as the set of sequences with finite support, i.e.,
\[
c_{00}=\left\{ (x_k)_{k\in\SN}\in\SR^\SN  \, \left|\, \sum_{k=1}^\infty 1\{x_k\neq 0\}<\infty\right.  \right\}. 
\]
Of course, $c_{00}$ is a dense subspace of $\ell_2$. Given the orthonormal basis $b_k$ and $\eta\in c_{00}$, we introduce the following perturbation to the density $f$: 
\begin{align}
f_{\eta}^{(T)}(e)=
f(e)\left(1+\frac{1}{\sqrt{T}}\sum_{k=1}^\infty \eta_k b_k(e)\right),\quad e\in\SR.\label{eqn:dens_perturb}
\end{align} 
The rate $T^{-1/2}$ is already indicative of the standard LAN behavior of the nuisance parameter $f$ as will formally follow from Proposition~\ref{prop:LAQ} below. 

For the symmetric case, we can assume that the perturbations $b_k$, $k\in\SN$, are also symmetric about zero.

The following proposition shows that the perturbations, both for the non-symmetric and for the symmetric case, are valid in the sense that they satisfy the conditions on the innovation density that we imposed throughout on the model (Assumption~\ref{ass:density}). The proof is organized in~\ref{supplementA}. 
\begin{proposition}\label{prop:dens_perturb}
Let $f\in\F$ and suppose $\eta\in c_{00}$. Then there exists $T^\prime\in\SN$ such that for all $T\geq T^\prime$ we have $f^{(T)}_\eta\in\F$. If we further restrict $f\in\Fs$ and $b_k$ is chosen symmetric about zero for $k\in\SN$, then
there exists $T^{\prime\prime}\in\SN$ such that for all $T\geq T^{\prime\prime}$ we have $f^{(T)}_\eta\in\Fs$.
\end{proposition}

\begin{remark}\label{remark:perturbations}
In semiparametric statistics one typically parametrizes perturbations to a density by a so-called ``non-parametric'' score function $h\in\rL_2^{0,f}$, i.e., the perturbation takes the form $f(e) k(T^{-1/2} h(e)))\approx f(e)(1+T^{-1/2} h(e))$ for a suitable function $k$; see, for example, \citet{BKRW} for details. By using the basis $b_k$, $k\in\SN$, we instead  tackle all such perturbations simultaneously via the infinite-dimensional nuisance parameter $\eta$. Of course, one would need to use $\ell_2$ as parameter space to ``generate'' all score functions $h$. We instead restrict to $c_{00}$ which ensures (\ref{eqn:dens_perturb}) to be a density (for large $T$). For our purposes this restriction will be without cost. Intuitively, this is since $c_{00}$ is a dense subspace of $\ell_2$ (so if a property is ``sufficiently continuous'' one only needs to establish it on $c_{00}$ because it  extends to the closure).
\end{remark}

\subsubsection*{Partial sum processes}
\noindent
To describe the limit experiment in Section~\ref{sec:limexp}, we introduce some partial sum processes and their limits. These results are fairly classical but, for completeness, precise statements are organized in Lemma~A.1 in the supplementary material.

Define, for $s\in [0,1]$, the partial sum processes
\begin{align*}
\WTe(s) =&~ \frac{1}{\sqrt{T}}\sum_{t=p+2}^{\lfloor sT\rfloor} \frac{\Gamma(L)\Delta Y_t}{\sigma_f},\\
\WTf(s) =&~ \frac{1}{\sqrt{T}}\sum_{t=p+2}^{\lfloor sT\rfloor} \sigma_f\phi_f( \Gamma(L)\Delta Y_t ), \\
\WTgamma(s) =&~ \frac{1}{\sqrt{T}}\sum_{t=p+2}^{\lfloor sT\rfloor} \left(\Delta Y_{t-1},\dots,\Delta Y_{t-p}\right)' \scoref(\Gamma(L)\Delta Y_t) \in \SR^{p},  \\
\WTbk(s) =&~ \frac{1}{\sqrt{T}}\sum_{t=p+2}^{\lfloor sT\rfloor} b_k( \Gamma(L)\Delta Y_t ), ~~~ k\in\SN.
\end{align*}
Note that we pick the starting point of the sums at $t=p+2$, so that these partial sum processes are (maximally) invariant with respect to the intercept $\mu$ (otherwise, e.g., for $t=p+1$, the term $\Gamma(L)\Delta Y_t$ contains $\Delta Y_1 = Y_1 - \mu$).

Using Assumption~\ref{ass:density} we find, under the null hypothesis, joint weak convergence of observation processes $\WTe$, $\WTf$, $\WTgamma$, and $\WTbk$ to Brownian motions that we denote by $\We$, $\Wf$, $\Wgamma$ and $\Wbk$, respectively.\footnote{All weak convergences in this paper are in product spaces of $D[0,1]$ with the uniform topology.} These limiting Brownian motions are defined on a probability space $(\Omega,\mathcal{F},\prob_{0,0,0})$. Let us already mention that we will introduce a collection of probability measures $\mathbb{P}_{h,\gamma,\eta}$, on $(\Omega, \mathcal{F})$, representing the limit experiment, in Section~\ref{sec:limexp}. We use the notational convention that probability measures related to the limit experiment (i.e., to the ``W-processes'') are denoted by $\mathbb{P}$, while probability measures related to the finite-sample unit root model, i.e., observing $Y_1, \dots ,Y_T$, will be denoted by $\law{}$.

We remark that integrals like $\int_0^1\We^{(T)}(s-) \rd\Wf^{(T)}(s)$ can be shown to converge weakly, under the null hypothesis, to the associated stochastic integral with the limiting Brownian motions, i.e., to $\int_0^1 \We(s)\rd\Wf(s)$. Weak convergence of integrals like $\int_0^1(\We^{(T)}(s-))^2 \rd s$ follows from an application of the continuous mapping theorem. Again, details can be found in the proof of Proposition~\ref{prop:LAQ} in the \ref{supplementA}.

\subsubsection*{\textbf{Behavior of $\We$, $\Wf$, $\Wgamma$ and $\Wbk$ under $\prob_{0,0,0}$ when $f\in\F$}}
\noindent
As $\varepsilon$ and $b_k(\varepsilon)$ are orthogonal for each $k$, it holds that $\We$ and $\Wbk$, $k\in\SN$, are all mutually independent. Moreover, we have
\begin{align*}
\var_{0,0,0} [\We(1)] = 1 {\rm~~~and~~}
\var_{0,0,0} [\Wbk(1)] = 1.
\end{align*}

As $\scoref(\ee)$ is the score of the location model, it is well known (see, for example, \citet{BKRW}) that we have (under Assumption~\ref{ass:density}) $\rE[f][\phi_f(\ee)]=0$ and $\rE[f][\ee\scoref(\ee)] =1$. Consequently, for $f\in \F$, again because $\ee$ and $b_k(\ee)$ are orthogonal for each $k$, we can decompose
\begin{align} \label{eqn:decomposition_score}
\sigma_f\phi_f(\ee)= \sigma_f^{-1}\varepsilon + \sum_{k=1}^\infty  
\FIaux{f,k} b_k(\varepsilon)
\end{align}
with coefficients $\FIaux{f,k}:=\sigma_f\rE[f][b_k(\varepsilon)\phi_f(\varepsilon)]$. This establishes
\begin{align}\label{eqn:decomposition_Wf}
\Wf = \We + \sum_{k=1}^\infty \FIaux{f,k} \Wbk.
\end{align}
Moreover, we have, for $k\in\SN$,
 \begin{align}\label{eqn:BMscore_covs}
\cov_{0,0,0}[\Wf(1), \We(1) ] = 1, ~~ 
\cov_{0,0,0}[\Wf(1), \Wbk(1)] = \FIaux{f,k}
\end{align}
and 
\begin{align}\label{eqn:decomposition_Jf}
\var_{0,0,0} [\Wf(1)]= \FIloc[f] = 1+\sum_{k=1}^\infty \FIaux{f,k}^2.
\end{align}

As for $\Wgamma$: since, under the null, $\left(\Delta Y_{t-1},\dots,\Delta Y_{t-p}\right)'$ is independent of the innovation $\ee_t$ and $\mathrm{E}_{0,0,0}\left[\Delta Y_{t-i}\right]=0$, $i=1,\dots,p$,  it follows 
\begin{align} \label{eqn:BMgamma1}
\cov_{0,0,0}[\Wgamma(1),\We(1)] = 0, ~~~
\cov_{0,0,0}[\Wgamma(1),\Wf(1)] = 0,
\end{align}
and
\begin{align} \label{eqn:BMgamma2}
\cov_{0,0,0}[\Wgamma(1),\Wbk(1)] = 0, ~~~k\in\SN.
\end{align}
We define the covariance matrix of $\Wgamma(1)$ as
\begin{align} \label{eqn:BMgamma3}
\var_{0,0,0} [\Wgamma(1)] = \Sigma_\Gamma \in \SR^{p\times p}
\end{align}
with
\begin{align*}
{\Sigma_\Gamma}_{i,j} := \FIloc[f]\rE\left[\Gamma(L)^{-1}\ee_{t-i}\Gamma(L)^{-1}\ee_{t-j}\right], ~~~ i,j = 1,\dots,p.
\end{align*}

\subsubsection*{\textbf{Behavior of $\We$, $\Wf$, $\Wgamma$ and $\Wbk$ under $\prob_{0,0,0}$ when $f\in\Fs$}}

 In this case, the density function $f(\ee)$ is an even function and so are the perturbation functions $b_k(\ee)$, $k\in\SN$. The score $\score_f(\ee)$ is now an odd function. Therefore, $\score_f(\ee)$ cannot be decomposed by $\ee$ and $b_k(\ee)$ anymore as in (\ref{eqn:decomposition_score}) and (\ref{eqn:decomposition_Wf}). Instead of (\ref{eqn:decomposition_Jf}) we now have 
\begin{align*}
\FIaux{f,k}=\sigma_f\rE[f][b_k(\varepsilon)\phi_f(\varepsilon)]=0 
\end{align*}
for all $f\in\Fs$ and $k\in\SN$. All the other results mentioned above still hold.

\subsection{A structural representation of the limit experiment}\label{sec:limexp}
\noindent
The results in the previous section are needed to study the asymptotic behavior of log-likelihood ratios. These in turn determine the limit experiment, which we use to study asymptotically optimal procedures invariant with respect to the nuisance parameters. Thus, fix $\mu\in\SR$, $\Gamma\in\Gspace$, and $f\in\F$. Let, for $h\in\SR$, $\gamma\in\SR^p$, and $\eta\in c_{00}$, $\law{h,\gamma,\eta;\mu,\Gamma,f}$ denote the law of $Y_1,\dots,Y_T$ under~(\ref{eqn:DGP1})-(\ref{eqn:DGP3}) with parameter $\rho$ given by~(\ref{eqn:local_alternative}), $\Gamma^{(T)}_\gamma(L)$ given by (\ref{eqn:local_gamma}), and innovation density~(\ref{eqn:dens_perturb}). The following proposition shows that the semiparametric unit root model is of the Locally Asymptotically Brownian Functional (LABF) type introduced in \citet{Jeganathan1995}.

\begin{proposition}\label{prop:LAQ}
Let $\mu\in\SR$, $\Gamma\in\Gspace$, $f\in \F$, $h\in\SR$, $\gamma\in\SR^p$, and $\eta\in c_{00}$. Let $\Delta$ denote differencing, i.e., $\Delta Y_t=Y_t-Y_{t-1}$.
\begin{enumerate}
\item[(i)] Then we have, under $\law{0,0,0;\mu,\Gamma,f}$,
\begin{align*}
\log \frac{\rd \law{h,\gamma,\eta;\mu,\Gamma,f}}{\rd \law{0,0,0;\mu,\Gamma,f}}
=&~ \sum_{t=1}^{T} \log \frac{ f_\eta^{(T)}\left(\Gamma^{(T)}_\gamma(L)\left(\Delta Y_t -\frac{h}{T} (Y_{t-1}-\mu)\right)\right)}{f\left(\Gamma(L)(\Delta Y_t)\right)} \\
=&~ h\CSrho + \gamma'\CS_\Gamma + \sum_{k=1}^\infty \eta_k \CSb{k} - \frac{1}{2}  \FIfin{}(h,\gamma,\eta) + o_P(1),\nonumber
\end{align*}
where the central-sequence $\CS = \big(\CSrho,\CS_\Gamma,\CSb{}\big)$, with $\CSb{} = \big(\CSb{k}\big)_{k\in\SN}$, is given by
\begin{align*}
\CSrho     & = \frac{1}{T}\sum_{t=p+2}^T  \Gamma(L)(Y_{t-1}-Y_1) \scoref(\Gamma(L)\Delta Y_t), \\
\CS_\Gamma & = \frac{1}{\sqrt{T}}\sum_{t=p+2}^{T}\left(\Delta Y_{t-1},\dots,\Delta Y_{t-p}\right)' \scoref(\Gamma(L)\Delta Y_t),   \\
\CSb{k}    & = \frac{1}{\sqrt{T}}\sum_{t=p+2}^T  b_k(\Gamma(L)\Delta Y_t), \quad k\in\SN,
\end{align*}
and 
\begin{align*}
\FIfin{}(h,\gamma,\eta)
=~& h^2 \FIloc \frac{1}{T^2}\sum_{t=p+2}^T \frac{\left(\Gamma(L)(Y_{t-1}-Y_1)\right)^2}{\sigma_f^2} + \gamma'\Sigma_\Gamma\gamma + \|\eta\|_2^2  \\
 & + 2h\frac{1}{T^{3/2}}\sum_{t=p+2}^T \frac{\Gamma(L)(Y_{t-1}-Y_1)}{\sigma_f} \sum_{k=1}^\infty \eta_k \FIaux{f,k}.
\end{align*}
\item[(ii)] Moreover, with $\CSlimrho=\int_0^1\We(s)\rd\Wf(s)$, $\CSlim_\Gamma = \Wgamma(1)$, and $\CSlimb{k}=W_{b_k}(1)$, $k\in\SN$, we have, still under  $\law{0,0,0;\mu,\Gamma,f}$  and as $T\to\infty$,
\begin{align} \label{eqn:LLRlimit}
\frac{\rd \law{h,\gamma,\eta;\mu,\Gamma,f}}{\rd \law{0,0,0;\mu,\Gamma,f}} 
\Rightarrow    
\exp\left(h \CSlimrho + \gamma'\CSlim_\Gamma + \sum_{k=1}^\infty \eta_k \CSlimb{k} -\frac{1}{2}\FIlim{}(h,\gamma,\eta) \right),
\end{align}
where 
\begin{align*}
\FIlim{}(h,\gamma,\eta)
=&~ h^2 \FIloc \int_0^1 (\We(s))^2 \rd s 
 + \gamma'\Sigma_\Gamma\gamma + \|\eta\|_2^2 \\
 &~ + 2 h \int_0^1
\We(s) \rd s
 \sum_{k=1}^\infty \eta_k  \FIaux{f,k}.  
\end{align*}
\item[(iii)] For all $h\in\SR$, $\gamma\in\SR^p$ and $\eta \in  c_{00}$ the right-hand side of~(\ref{eqn:LLRlimit}) has unit expectation under $\prob_{0,0,0}$.
\end{enumerate}
\end{proposition}
\noindent
Of course, Proposition~\ref{prop:LAQ} still holds for $f\in\Fs$; in that case we have $\FIaux{f,k}=0$.  The proof of~(i) follows by an application of Proposition~1 in  \citet{HvdAW2} which provides generally applicable sufficient conditions for the quadratic expansion of log likelihood ratios. Of course, Part~(ii) is not surprising and follows using the weak convergence of the partial sum processes to Brownian motions (and integrals involving the partial sum processes to stochastic integrals) discussed above. Finally, Part~(iii) follows by verifying the Novikov condition. For the sake of completeness, detailed proofs are organized in \ref{supplementA}. 

Part~(iii) of the proposition implies that we can introduce, for $h\in\SR$, $\gamma\in\SR^p$, and $\eta\in c_{00}$, new probability measures $\prob_{h,\gamma,\eta}$ on the measurable space $(\Omega,\mathcal{F})$ (on which the processes $\We$, $\Wf$, $\Wgamma$ and $\Wbk$ were defined) by their Radon-Nikodym derivatives with respect to $\prob_{0,0,0}$:
\begin{align*}
\frac{\rd \prob_{h,\gamma,\eta}}{\rd \prob_{0,0,0}}=\exp\left(h\CSlimrho + \gamma'\CSlim_\Gamma + \sum_{k=1}^\infty \eta_k \CSlimb{k} 
-\frac{1}{2}  \FIlim{}(h,\gamma,\eta)\right).
\end{align*}
Proposition~\ref{prop:LAQ} then implies that the sequence of (local) unit root experiments (each $T\in\SN$ yields an experiment) weakly converges (in the Le Cam sense) to the experiment described by the probability measures $\prob_{h,\gamma,\eta}$. Formally, we define the sequence of experiments of interest by 
\begin{align*}
\limitET(\mu,\Gamma,f)=\left( \SR^T, \mathcal{B}(\SR^T), ( \law{h,\gamma,\eta;\mu,\Gamma,f}\, |\, h\in\SR,\, \gamma\in\SR^p, \eta\in c_{00})\right)
\end{align*}
for $T\in\SN$, and the limit experiment by, with $\mathcal{B}_C$ the Borel $\sigma$-field on $C[0,1]$, 
\begin{align*}
\limitE(f) = &~ \Big( C[0,1] \times C[0,1] \times C^{p}[0,1] \times C^{\SN}[0,1], \\
&~~~~ \mathcal{B_C} \otimes \mathcal{B_C} \otimes (\otimes_{p} \mathcal{B_C}) \otimes (\otimes_{k=1}^\infty \mathcal{B_C}), \left(\prob_{h,\gamma,\eta}\,|\, h\in\SR, \gamma\in\SR^p, \eta\in c_{00}\right)\Big).
\end{align*}
Note that the latter experiment indeed depends on $f$ as the measure $\prob_{h,\gamma,\eta}$ depends on $f$.

\begin{corollary}\label{cor:convexp}
Let $\mu\in\SR$, $\Gamma\in\Gspace$, and $f\in\F$. Then the sequence of experiments $\limitET(\mu,\Gamma,f)$, $T\in\SN$, converges to the experiment $\limitE(f)$ as $T\to\infty$.
\end{corollary}
\noindent
The Asymptotic Representation Theorem (see, for example, Chapter~9 in \citet{vdVaart00}) implies that for any statistic $A_T$ which converges in distribution to the law $L_{h,\gamma,\eta}$, under $\law{h,\gamma,\eta;\mu,\Gamma,f}$, there exists a (possibly randomized) statistic $A$, defined on $\limitE(f)$, such that the law of $A$ under $\prob_{h,\gamma,\eta}$ is given by $L_{h,\gamma,\eta}$. This allows us to study (asymptotically) optimal inference: the ``best'' procedure in the limit experiment also yields a bound for the sequence of experiments. If one is able to construct a statistic (for the sequence) that attains this bound, it follows that the bound is sharp and the statistic is called (asymptotically) optimal. This is precisely what we do: Section~\ref{sec:invariance} establishes the bound and in Section~\ref{sec:tests} we introduce statistics attaining it.

To obtain more insight in the limit experiment $\limitE(f)$
the following proposition, which follows by a direct application of Girsanov's theorem, provides a ``structural'' description of the limit experiment. 

\begin{proposition}\label{prop:Girsanov}
Let $h\in\SR$, $\gamma\in\SR^p$, $\eta\in c_{00}$, and $f \in \F$. The processes $\Ze$, $\Zf$, $\Zgamma$ and $Z_{b_k}$, $k\in\SN$, defined by the starting values $\Ze(0)=\Zf(0)=\Zbk(0)=0$, $\Zgamma(0) = \boldsymbol{0}_{p}$, and the stochastic differential equations, for $s\in[0,1]$,
\begin{align*}
\rd \Ze(s)  &= \rd \We(s)  - h \We(s)\rd s, \\
\rd \Zf(s)  &= \rd \Wf(s)  - h \FIloc \We(s)\rd s - \textstyle\sum_{k}\eta_k\FIaux{f,k}\rd s, \\
\rd \Zgamma(s) &= \rd \Wgamma(s) - \gamma \rd s, \\
\rd \Zbk(s) &= \rd \Wbk(s) - h \FIaux{f,k} \We(s)\rd s - \eta_k\rd s, \quad k\in\SN,
\end{align*}
are zero-drift Brownian motions under $\prob_{h,\gamma,\eta}$. Their joint law is that of $(\We,\Wf,\Wgamma,(\Wbk)_{k\in\SN})$ under $\prob_{0,0,0}$.

\end{proposition}

For the case $f\in\Fs$, Proposition~\ref{prop:Girsanov} still applies with $\FIaux{f,k}=0$. Moreover, for this case, we denote by $\limitETs(f)$ the associated sequence of experiments and by $\limitEs(f)$ the associated limit experiment. 

\begin{remark} \label{remark:mu}
Part (i) and (ii) Proposition~\ref{prop:LAQ} show that the parameter $\mu$ vanishes in the limit. More explicitly, in the proof of this proposition, we replace $\mu$ in the likelihood ratio term by $Y_1$ and then show that the difference term is $o_P(1)$. On the other hand, one could also ``localize'' the parameter $\mu$ as $\mu=\mu^{(T)}_d=\mu_0+d$ (with rate $T^0=1$) as in \citet{Jansson2008}. As shown in that paper, the term associated to the parameter $d$ does not change with $T$ and is independent of the other terms of the likelihood ratio. By the additively separable structure, we can treat the parameter $\mu$ ``as if'' it is known. In either way, inference for $\rho$ would be invariant with respect to $\mu$ in the limit. Analogously, in the finite-sample experiment $\limitET(f)$, $\mu$ is eliminated (automatically) by using the increments $\Delta Y_t$, $t=2,\dots,T$, which are (maximally) invariant with respect to $\mu$ (see Section~\ref{sec:tests}). 
\end{remark}
\subsection{The limit experiment: invariance and power envelope}\label{sec:invariance}
\noindent
In this section, we consider the limit experiments $\limitE(f)$ and $\limitEs(f)$. In these experiments, we observe the processes $\We$, $\Wf$, $\Wgamma$, and $W_{b_k}$, $k\in\SN$, continuously on the time interval $[0,1]$ from the model $(\prob_{h,\gamma,\eta}\, |\, h\in\SR,\,\gamma\in\SR^p,\,\eta\in c_{00})$, and we are interested in the power envelopes for testing the hypothesis
\begin{equation}\label{eqn:hypo_limexp}
\rH_0:\, h=0 ~ (\gamma\in\SR^p, \eta\in c_{00}) \text{ versus }
	\rH_a:\, h<0 ~ (\gamma\in\SR^p, \eta\in c_{00})
\end{equation}
To eliminate the nuisance parameters $\gamma$ and $\eta$, we first propose a statistic that is sufficient for the parameter of interest $h$ and does not depend on $\gamma$. Afterwards, using Proposition~\ref{prop:Girsanov}, we discuss a natural invariance structure with respect to the infinite-dimensional nuisance parameter $\eta$. We derive the maximal invariant and apply the Neyman-Pearson lemma to obtain the power envelopes of invariant tests within the experiments $\limitE(f)$ and $\limitEs(f)$, respectively in Section~\ref{subsubsec:eliminate_eta_Ef} and Section~\ref{subsubsec:eliminate_eta_ESf}.

We begin with the elimination of $\gamma$. The statistic $\left(\We,\Wf,(\Wbk)_{k\in\SN}\right)$ serves as a sufficient statistic for the parameter $h$. This is because, according to the structural version of $\limitE(f)$ (or $\limitEs(f)$) in Proposition~\ref{prop:Girsanov}, the distribution of $\Wgamma$ is only affected by $\gamma$ and so is the distribution of the process $\left(\We,\Wf,(\Wbk)_{k\in\SN}\right)$ only affected by $h$ and $\eta$. It then follows that the distribution of the statistic $\left(\We,\Wf,\Wgamma,(\Wbk)_{k\in\SN}\right)$ conditional on the statistic $\left(\We,\Wf,(\Wbk)_{k\in\SN}\right)$ is only a function of $\Wgamma$ and $\gamma$ and, in turn, does not depend on $h$. Next, observe that the sufficient statistic $\left(\We,\Wf,(\Wbk)_{k\in\SN}\right)$ is independent of the process $\Wgamma$ and thus has a distribution that does not depend on $\gamma$. This allows us to restrict attention to the sufficient statistic $\left(\We,\Wf,(\Wbk)_{k\in\SN}\right)$ to conduct inference for $h$, and the nuisance parameter $\gamma$ disappears from the likelihoods.


\subsubsection{Elimination of $\eta$ in $\limitE(f)$ and the associated power envelope} \label{subsubsec:eliminate_eta_Ef}
The elimination of the nuisance parameter $\eta$ is more involved and different for the limit experiments $\limitE(f)$ and $\limitEs(f)$. We start with $\limitE(f)$.

In Proposition~\ref{prop:Girsanov}, if one applies the decompositions (\ref{eqn:decomposition_Wf}) and (\ref{eqn:decomposition_Jf}) to the first equation (of $\We$) and the fourth equation (of $\Wbk$), one retrieves the second equation (of $\Wf$). This essentially allows us to omit the process $\Wf$ and restrict the observations to the processes $\We$ and $\Wbk$, $k\in\SN$. 

Now we formalize the invariance structure with respect to $\eta$. Introduce, for $\eta\in c_{00}$, the transformation $g_\eta=(g_{\eta_k})_{k\in\SN}: C^\SN[0,1]\to C^\SN[0,1]$ defined by, for $W\in C[0,1]$,
\begin{equation}\label{eqn:DefinitionGroupG}
g_{\eta_k}:\,\,\left[g_{\eta_k}(W)\right](s)=W(s)-\eta_k s,~~s\in[0,1],
\end{equation}
i.e., $g_{\eta_k}$ adds a drift $s\mapsto -\eta_k s$ to its argument process. 
Proposition~\ref{prop:Girsanov} implies that the law of $(\We,(g_{\eta_k}(W_{b_k}))_{k\in\SN})$ under $\mathbb{P}_{h,\gamma,0}$ is the same as the law of  $(\We,(W_{b_k})_{k\in\SN})$ under $\mathbb{P}_{h,\gamma,\eta}$. 
Hence our testing problem~(\ref{eqn:hypo_limexp}) is invariant with respect to the transformations $g_\eta$. Therefore, following the invariance principle, it is natural to restrict attention to test statistics that are invariant with respect to these transformations as well, i.e., test statistics $t$ that satisfy
\begin{equation}\label{eqn:test_invariance}
t(\We,(g_{\eta_k}(W_{b_k}))_{k\in\SN})= t(\We,(W_{b_k})_{k\in\SN}) \text{ for all } g_\eta,\, \eta\in c_{00}.
\end{equation}
Given a process $W$ let us define the associated \textit{bridge process} by $B^W(s)=W(s)- s W(1)$. Now note that we have, for all $s\in [0,1]$ and $k\in\SN$,
\begin{align*}
B^{g_{\eta_k}(W)}(s)&= [g_{\eta_k}(W)](s) - s [g_{\eta_k}(W)](1)\\
&=W(s)- s \eta_k - s ( W(1) - 1\times \eta_k)\\
&=W(s)- s W(1) \\
&=B^{W}(s),	
\end{align*}
i.e., taking the bridge of a process ensures invariance with respect to adding drifts to that process. Define the mapping $M$ by 
\begin{align*}
M(\We, (\Wbk)_{k\in\SN}):=(\We, (\Bbk)_{k\in\SN}),
\end{align*}
where $\Bbk := B^{\Wbk}$. It follows that statistics that are measurable with respect to the $\sigma$-field,
\begin{align}\label{eqn:maximal_invariant}
\maxInvf = \sigma\left(M(\We, (\Wbk)_{k\in\SN})\right) = \sigma(\We, (\Bbk)_{k\in\SN}),
\end{align}
are invariant (with respect to $g_\eta$, $\eta\in c_{00}$). It is, however, not (immediately) clear that we did not throw away too much information. Formally, we need $\maxInvf$ to be \textit{maximally invariant} which means that each invariant statistic is $\maxInvf$-measurable. The following theorem, which once more exploits the structural description of the limit experiment, shows that this indeed is the case.
\begin{theorem}\label{thm:invariance}
The $\sigma$-field $\maxInvf$ in~(\ref{eqn:maximal_invariant}) is maximally invariant for the group of transformations $g_\eta$, $\eta\in c_{00}$, in the experiment $\limitE(f)$. 
\end{theorem}
The above theorem implies that invariant inference must be based on $\maxInvf$. An application of the Neyman-Pearson lemma, using $\maxInvf$ as observation, yields the power envelope for the class of invariant tests. To be precise, consider the likelihood ratios restricted to $\maxInvf$, which are given by
\begin{align*}
\frac{\rd\prob_{h}^{\maxInvf}}{\rd\prob_{0}^{\maxInvf}} = \mathbb{E}_0 
\left[ \frac{\rd \prob_{h,\gamma,\eta} }{\rd \prob_{0,\gamma,\eta}}  
\mid  \maxInvf \right], 	
\end{align*}
where the conditional expectation indeed does not depend on $\eta$ precisely because of the invariance. The conditional expectation also does not depend on $\gamma$ as a result of the arguments stated at the beginning of this section.

To calculate this conditional expectation, we first introduce $\Bf=B^{W_{\scoref}}$, i.e., the bridge process associated to $\Wf$. Following the decomposition in (\ref{eqn:decomposition_Wf}), we can decompose $\CSlimrho=\int_0^1\We(s)\rd\Wf(s)=I+II$ with
\begin{align*}
I &=
\int_0^1\We(s)\rd\Bf(s) + \We(1)\int_0^1\We(s) \rd s,\\
II &=
\left( \sum_{k=1}^\infty \FIaux{f,k}\Wbk(1) \right) \int_0^1\We(s)\rd s. 
\end{align*}
Note that part $I$ is $\maxInvf$-measurable. Under $\prob_{0,0,0}$ the random variables $\Wbk(1)$, $k\in\SN$, are independent of $\We$ and $\Bbk$, $k\in\SN$. Indeed, the independence of $\We$ holds by construction and the independence of $B_{b_k}$ is a well-known, and easy to verify, property of Brownian bridges. We thus obtain, since $\FIlim{}(h,\gamma,\eta)$ is $\maxInvf$-measurable as well,
\begin{align*}
\frac{\rd \prob_{h}^{\maxInvf} } {\rd \prob_{0}^{\maxInvf}}
=&~
\mathbb{E}_0\left[ \frac{\rd \prob_{h,\gamma,\eta} }{\rd \prob_{0,\gamma,\eta}} \mid \maxInvf\right] \\
=&~
\exp\left( h\times I -\frac{1}{2} \FIlim{}(h,\gamma,\eta) \right)\\
 & \mbox{}\times \mathbb{E}_{0,0,0} \left[\exp\left(\sum_{k=1}^\infty ( h \FIaux{f,k} \int_0^1 \We(s) \rd s +\eta_k)\Wbk(1) + \gamma^\prime\Wgamma(1)
\right) \mid \maxInvf \right] \\
=&~
\exp\Bigg( h\times I - \frac{1}{2}\left( \FIlim{}(h,\gamma,\eta) - \sum_{k=1}^\infty ( h \FIaux{f,k}  \int_0^1 \We(s) \rd s +\eta_k)^2 - \gamma^\prime\Sigma_\Gamma\gamma \right)\Bigg) \\
=&~
\exp\left( h \EffCS_f - \frac{1}{2} h^2 \EffFI_f\right)
\end{align*}
with
\begin{align}
\label{eqn:EfficientCentralSequence}
\EffCS_f 
= &~ \int_0^1 \We(s) \rd \Bf(s) + \We(1) \int_0^1 \We(s) \rd s,\\
\label{eqn:EfficientFisherInformation}
\EffFI_f 
= &~ J_{f}\int_0^1\We^2(s) \rd s-\left(\int_0^1\We(s)\rd s\right)^2\sum_{k=1}^\infty \FIaux{f,k}^2\\\nonumber
= &~ J_f\int_0^1\We^2(s) \rd s-\left(\int_0^1\We(s)\rd s\right)^2\left(J_{f}-1\right),
\end{align}
where the last equality follows from~(\ref{eqn:decomposition_Jf}). Note that the distribution of this likelihood ratio indeed does not depend on the nuisance parameters $\gamma$ or $\eta$. 

We can now formalize the notion of point-optimal invariant tests in the limit experiment. To that end, let us denote the $(1-\alpha)$-quantile of $\rd\prob_{h}^{\maxInvf}/\rd\prob_{0}^{\maxInvf}$ under $\mathbb{P}_{0,\gamma,\eta}$, which does not depend on either $\gamma$ or $\eta$, by $c(h,J_f;\alpha)$. Define the size-$\alpha$ test $\phi^{\F*}_{f,\alpha}(\bar{h}) = \indicator{\rd\prob_{\bar{h}}^{\maxInvf} / \rd \prob_{0}^{\maxInvf}\geq c(\bar{h},J_f;\alpha)}$, for a fixed value of $\bar{h}<0$. Note that this is an oracle test in $\mathcal{E}(f)$ that depends on the true value of $f$. A feasible test is provided in Section~\ref{sec:tests}. The power function of this oracle test is given by
\begin{align*}
h \mapsto \pi^{\F*}_{f,\alpha}(h; \bar{h}) = \mathbb{E}_0\left[\phi^{\F*}_{f,\alpha}(\bar{h}) \frac{\rd\prob_{h}^{\maxInvf}}{\rd \prob_{0}^{\maxInvf}}\right] = \mathbb{E}_0\left[\phi^{\F*}_{f,\alpha}(\bar{h}) \frac{\rd\prob_{h,\gamma,\eta}}{\rd \prob_{0,0,0}}\right].
\end{align*}
An application of the Neyman-Pearson lemma yields the following.	

\begin{corollary}\label{cor:power_envelope_in_limit} 
Let $f\in\F$ and $\alpha\in (0,1)$. Let $\phi$ be a (possibly randomized) test that is $\maxInvf$-measurable and is of size $\alpha$, i.e., $\mathbb{E}_{0} \phi \leq \alpha$. Let $\pi$ denote the power function of this test, i.e., $\pi(h)=\mathbb{E}_{h} \phi$. Then we have 
\begin{align*}
\pi(\bar{h})\leq \pi^{\F*}_{f,\alpha}(\bar{h};\bar{h}), ~~~ \bar{h}<0. 
\end{align*} 
\end{corollary}
\noindent
The (oracle) test $\phi^{\F*}_{f,\alpha}(\bar{h})$ thus is point optimal, i.e., its power function is tangent to the (semiparametric) power envelope $h\mapsto \pi^{\F*}_{f,\alpha}(h; h)$ at $h=\bar{h}$.\footnote{Here and later in this section, the early usage of the concept ``power envelope'' is due to the fact that it is shown to be the upper bound later in this section and point-wisely attainable by tests in sequence in Section~\ref{sec:tests}.}

\begin{remark} \label{remark:ERSlimit}
The notion of invariance in the limit experiment leads to another interpretation of the \citet{ERS1996} (ERS) test statistic. Note that $\sigma$-field $\Inve=\sigma\left(\We(s);\, s\in [0,1]\right)$ is also invariant, though not maximally so. The likelihood ratio conditional on observing only $\Inve$ is given by

\begin{align*}
\frac{\rd \prob_{h}^{\Inve}}{\rd \prob_{0}^{\Inve}}
 =&~
\mathbb{E}_0 
\left[ 
\frac{\rd \prob_{h}^\maxInvf } {\rd \prob_{0}^\maxInvf}
\mid \Inve 
\right]\\
 =&~
\exp\left(
	h\int_0^1\We(s) \rd \Be(s)+ h \We(1) \int_0^1 \We(s) \rd s -\frac{1}{2} h^2 \EffFI\right)\\
  &~
\mbox{}\times\mathbb{E}_0 
\left[ 
\exp\left(h\int_0^1\We(s)\rd B_b(s)\right)
\mid  \Inve 
\right]\\
 =&~
 \exp\left(
 h\int_0^1\We(s) \rd \We(s) - \frac{1}{2} h^2 \EffFI\right)\\
  &~
\mbox{}\times
\exp\left(\frac12h^2\left[\int_0^1\We^2(s)\rd s - \left(\int_0^1\We(s)\rd s\right)^2\right]\left(J_f-1\right)\right)\\
 =&~
 \exp\left(
 h\int_0^1 \We(s) \rd \We(s)
 -\frac{1}{2}h^2\int_0^1\We^2(s)\rd s\right),
\end{align*}
where $\Wb(s) = \sum_{k=1}^\infty \FIaux{f,k}\Wbk(s) $ and $\Bb(s) = B^{\Wb}(s)$ for notational simplicity. 
As a result, the ERS test statistic equals the likelihood ratio statistic using the (non-maximal) invariant $\Inve$. This explains the improved power of our tests within the model we consider. Moreover, for Gaussian $f$, we have $\Inve=\maxInvf$ and obtain point-optimality of the ERS test.\footnote{Similarly, one could try to derive the statistic resulting from using $\mathcal{M}_{B}=\sigma\left(B_{b_k}(s);\, s\in [0,1]\right)$ as an invariant. However, that does not seem to lead to an insightful result.}
\end{remark}

\begin{remark} \label{remark:JanssonPowerEnvelope}
The semiparametric power envelope derived above for this case of $f\in\F$, of course, coincides with the one in \citet{Jansson2008} based on the invariance constraint. This can be seen by rewriting $\EffCS_f=\int_0^1 \We(s) \rd \Wf(s) - (\Wf(1) - \We(1)) \int_0^1 \We(s) \rd s$. We feel our approach is attractive since, by describing the perturbations on $f$ with an orthonormal basis and a infinite-dimensional parameter instead of one single parameter, there is no need to find the least favorable direction.\footnote{The traditional semiparametric approach, developed for LAN-type experiments, accomplishes this by projecting the score function of the parameter of interest onto the tangent space of nuisance score functions. However, this approach seems not easily generalizable to LABF-type experiments.} Moreover, we feel that the use of the invariance principle, rather than the similarity constraint, more naturally suggests (partly) rank-based tests.
\end{remark}

\subsubsection{Elimination of $\eta$ in $\limitEs(f)$ and the associated power envelope} \label{subsubsec:eliminate_eta_ESf}

Since in this case we have $\FIaux{f,k}=0$, the structural representation of experiment $\limitEs(f)$ in Proposition~\ref{prop:Girsanov} becomes 
\begin{align*}
\rd Z_\varepsilon(s) &= \rd \We(s) - h \We(s)\rd s, \\
\rd \Zf(s) &= \rd \Wf(s) - h J_f\We(s)\rd s, \\
\rd Z_{b_k}(s) &= \rd W_{b_k}(s) - \eta_k\rd s, \quad k\in\SN.
\end{align*}
Note that here the process $\Wgamma$ is again removed by sufficiency in order to eliminate the nuisance parameter $\gamma$ (see the discussion at the beginning of Section~\ref{sec:invariance}). Following the same argument for the case of $f\in\F$, statistics that are measurable with respect to the $\sigma$-field
\begin{align}\label{eqn:maximal_invariant_symmetric}
\maxInvfs = \sigma\left(M(\We, \Wf, (\Wbk)_{k\in\SN})\right) = \sigma(\We, \Wf, (\Bbk)_{k\in\SN})
\end{align}
are invariant with respect to the transformations $g_\eta$, $\eta\in c_{00}$.
Moreover, in the following theorem, we show that this $\sigma$-field is maximally invariant. 

\begin{theorem}\label{thm:invariance_symmetric}
The $\sigma$-field $\maxInvfs$ in~(\ref{eqn:maximal_invariant}) is maximally invariant for the group of transformations $g_\eta$, $\eta\in c_{00}$, in the experiment $\limitEs(f)$. 
\end{theorem}

The likelihood ratio restricted to $\maxInvfs$ is given by
\begin{align*}
\frac{\rd \prob_{h}^{\maxInvfs} } {\rd \prob_{0}^{\maxInvfs}}
=&~
\mathbb{E}_0\left[ \frac{\rd \prob_{h,\gamma,\eta} }{\rd \prob_{0,\gamma,\eta}} \mid \maxInvfs\right] \\
=&~
\exp\left( h \CSlim_f -\frac{1}{2} \FIlim{}(h,\gamma,\eta) \right)\\
 & \mbox{}\times \mathbb{E}_{0,0,0} \left[\exp\left(\sum_{k=1}^\infty \eta_k\Wbk(1) + \gamma^\prime\Wgamma(1) \right) \mid \maxInvfs \right] \\
=&~
\exp\left( h\CSlim_f -\frac{1}{2} \FIlim{}(h,\gamma,\eta) + \frac{1}{2}\|\eta\|_2^2 + \frac{1}{2}\gamma^\prime\Sigma_\Gamma\gamma \right) \\
=&~
\exp\left( h \CSlim_f - \frac{1}{2} h^2 \FIlim_f\right),
\end{align*}
where 
\begin{align} \label{eqn:EfficientCSFI_symmetric}
\FIlim_f := J_f\int_0^1\We^2(s) \rd s.
\end{align}

As before, we then formalize the point-optimal invariant tests in $\limitEs(f)$. Denote the $(1-\alpha)$-quantile of $\rd\prob_{h}^{\maxInvfs}/\rd\prob_{0}^{\maxInvfs}$ under $\mathbb{P}_{0,\gamma,\eta}$ by $c_{\mathbb{S}}(h,J_f;\alpha)$. Define the size-$\alpha$ test $\phi^{\Fs*}_{f,\alpha}(\bar{h}) = \indicator{\rd\prob_{\bar{h}}^{\maxInvfs} / \rd \prob_{0}^{\maxInvfs}\geq c_{\mathbb{S}}(\bar{h},J_f;\alpha)}$, for a fixed value of $\bar{h}<0$. The associated power function is
\begin{align*}
h\mapsto \pi^{\Fs*}_{f,\alpha}(h; \bar{h}) = \mathbb{E}_0\left[\phi^{\Fs*}_{f,\alpha}(\bar{h}) \frac{\rd\prob_{h,\gamma,\eta}}{\rd \prob_{0,0,0}}\right].
\end{align*}
Again, by the Neyman-Pearson lemma, the following corollary holds.

\begin{corollary}\label{cor:power_envelope_in_limit_symmetric} 
Let $f\in\Fs$ and $\alpha\in (0,1)$. Let $\phi$ be a (possibly randomized) test that is $\maxInvfs$-measurable and is of size $\alpha$, i.e., $\mathbb{E}_{0} \phi \leq \alpha$. Let $\pi$ denote the power function of this test, i.e., $\pi(h)=\mathbb{E}_{h} \phi$. Then we have 
\begin{align*}
\pi(\bar{h}) \leq \pi^{\Fs*}_{f,\alpha}(\bar{h};\bar{h}). 
\end{align*} 
\end{corollary}

The likelihood ratio of the maximal invariant $\maxInvfs$, $\rd \prob_{h}^{\maxInvfs} / \rd \prob_{0}^{\maxInvfs}$, equals the likelihood ratio of the full observation $(\We,\Wf,\Wgamma,(\Wbk)_{k\in\SN})^\prime$ with known $\eta=0$ (i.e., known $f$) and $\gamma=0$,
$\rd \prob_{h,0,0} / \rd \prob_{0,0,0}$. This shows that the ``semiparametric'' power envelope $\pi^{\Fs*}_{f,\alpha}$ actually coincides with the parametric power envelope, which is defined based on the likelihood ratio $\rd \prob_{h,0,0} / \rd \prob_{0,0,0}$. This verifies again the adaptivity result in \cite{Jansson2008} under the same conditions for this unit root testing problem.\footnote{A discussion about the notion of ``adaptiveness'' in this nonstandard testing problem can be found in Section 5 of \cite{Jansson2008}.} We provide, in Sections \ref{sec:hybrid_rank_tests} and \ref{sec:approximatehybridranktests}, a class of adaptive unit root tests based on signed-rank statistics for this setting.

\begin{remark} \label{remark:scaleinvariant}
The semiparametric power envelopes $\pi^{\F*}_{f,\alpha}$ and $\pi^{\Fs*}_{f,\alpha}$ are scale invariant, i.e., invariant with respect to the value of $\sigma_f>0$. This is easily seen from the fact that $\We$, $\Wf$ and $J_f$ are all scale invariant.  
\end{remark}

\begin{remark} \label{remark:alternative_elimination}
The problem of eliminating the nuisance parameter $\eta$ in the symmetric density case ($f\in\Fs$) is actually the same as that of eliminating the autocorrelation parameter $\gamma$ in the beginning of this section: the nuisance parameter appears only in a process which is independent of all the other processes. To be specific, $\eta$ only affects the distribution of $\Wbk$, which is independent of $\Wf$ as well as $\We$. Therefore, the distribution of $\left(\We,\Wf,(\Wbk)_{k\in\SN}\right)$ conditional on the statistic $\left(\We,\Wf\right)$ does not depend on the parameter of interest $h$. Hence, the statistic $\left(\We,\Wf\right)$ serves as a sufficient statistic for $h$ which is also invariant with respect to $\eta$. 
\end{remark}

\subsection{The asymptotic power envelope for asymptotically invariant tests }\label{sec:ART}
\noindent
Now we translate the results for the limit LABF experiment to the unit root model of interest. To mimick the invariance in the limit experiment we introduce the following definition.

\begin{definition}
Let $\mu\in\SR$, $\Gamma\in\Gspace$, $f\in\F$. A sequence of test statistics $\psi^{(T)}$ is said to be asymptotically invariant if the distribution of $\psi^{(T)}$ weakly converges, under $\law{h,\gamma,\eta;\mu,\Gamma,f}$ for all $h\leq 0$, $\gamma\in\SR^p$, and $\eta\in c_{00}$, to the distribution of an invariant test in the limit experiment $\limitE(f)$/$\limitEs(f)$, under $\prob_{h,\gamma,\eta}$.
\end{definition}

The Asymptotic Representation Theorem (see, e.g., \citet{vdVaart00} Chapter~9) now yields the following main result on the asymptotic power envelope. 

\begin{theorem}\label{thm:asymptotic_power_envelope}
Let $\mu\in\SR$, $\Gamma\in\Gspace$, $f\in\F$, and $\alpha\in (0,1)$. 
Let $\phi_T(Y_1,\dots,Y_T)$, $T\in\SN$, be an asymptotically invariant test of size $\alpha$, i.e., $\limsup_{T\to\infty} \mathrm{E}_{0,\gamma,\eta} \phi_T\leq \alpha$ for all $\gamma\in\SR^p$ and $\eta\in c_{00}$. 
Let $\pi_T$ denote the power function of $\phi_T$, i.e., $\pi_T(h,\gamma,\eta)=\mathrm{E}_{h,\gamma,\eta;\mu,\Gamma,f} \phi_T$. Then, we have 
\begin{align*}
\limsup_{T\to\infty} \pi_T(h,\gamma,\eta)\leq \pi^{\F*}_{f,\alpha}(h ; h),\quad h<0,~ \gamma\in\SR^p, {\rm~and~} \eta\in c_{00}.
\end{align*}
For the case of $f\in\Fs$, we have 
\begin{align*}
\limsup_{T\to\infty} \pi_T(h,\gamma,\eta)\leq \pi^{\Fs*}_{f,\alpha}(h ; h),\quad h<0,~ \gamma\in\SR^p, {\rm~and~} \eta\in c_{00}.
\end{align*}
\end{theorem}
These power envelopes for invariant tests in the limit experiments $\limitE(f)$ and $\limitEs(f)$ thus provide upper bounds to the asymptotic powers of invariant tests for the unit root hypotheses in $\limitET(f)$ and $\limitETs(f)$, respectively. The next section introduces two classes of tests (based on rank and signed-rank statistics) that (in a point-wise sense) attain these power envelopes and, thereby, demonstrates that these bounds are sharp. We additionally provide a Chernoff-Savage type result for these classes of tests.

\section{A class of semiparametrically optimal hybrid rank tests}\label{sec:tests}
\noindent The appearance of the bridge process $\Bf$ in the ``efficient central sequence'' $\Delta^*_f$ naturally suggests the (partial) use of ranks in the construction of test statistics. Indeed, we can construct an empirical analogue of $B_{\scoref}$ by considering a partial-sum process which only depends on the observations via the ranks $R_t$ of $\Gamma(L)\Delta Y_t$ amongst $\Gamma(L)\Delta Y_{p+2},\dots, \Gamma(L)\Delta Y_T$. We allow for the use of a \emph{reference density} $g$ that may or may not be equal to the true underlying innovation density $f$. Our findings compare to Quasi-ML methods: if the true innovation density happens to be the same as the selected reference density the inference procedure is point-optimal. At the same time, the procedure remains valid, i.e., has proper asymptotic size, even in case the true innovation density does not coincide with the reference density. Note that these results also hold in case the reference density is non-Gaussian, while Quasi-ML results are generally restricted to Gaussian reference densities.

We need the following mild assumption on the reference density.
\begin{assumption}\label{ass:ref_density}
The density $g\in\F$, with finite variance $\sigma_g^2$, satisfies
\begin{equation*}
\lim_{T\rightarrow\infty}\frac{1}{T}\sum_{i=1}^T \sigma_g^2\score_g^2\left(G^{-1}\left(\frac{i}{T+1}\right)\right) = J_g,
\end{equation*}
with location score function $\score_g(\ee):=-(g^\prime/g)(\ee)$, where $J_g$ is the standardized Fisher information for location of $g$.\footnote{Similarly to the standardized Fisher information $J_f$ of $f$, the Fisher information $J_g$ of $g$ is standardized by the variance $\sigma_g^2$. As a result, it is scale invariant.}
\end{assumption}

Now we can formulate the following direct extension of Lemma~A.1 in \citet{HvdAW1} and its signed-rank counterpart. The proof for the weak convergence of the stochastic integrals in (\ref{eqn:stochasticintegralconvergence_Bg}) and (\ref{eqn:stochasticintegralconvergence_Wg}) is provided in \ref{supplementA}. 
\begin{lemma}\label{lem:rankprocess}
Let $\mu\in\SR$, $\Gamma\in\Gspace$, and $g$ satisfy Assumption~\ref{ass:ref_density}.
\begin{itemize}
\item[(i)] For the case $f\in\F$, consider the partial sum process
\begin{equation}\label{eqn:partsum_Bg}
\BTg(s) =
\frac{1}{\sqrt{T}}\sum_{t=p+2}^{\lfloor s T \rfloor} \sigma_g \left(\phi_g\left( G^{-1}\left(\frac{R_t}{T-p}\right)\right) - \bar{\phi}_g^{(T)}\right)
\end{equation}
for $s\in[0,1]$, where $\bar{\phi}_g^{(T)} := T^{-1}\sum_{i=1}^{T-p-1}\phi_g(G^{-1}(i/(T-p)))$, and $R_t$ denotes the rank of $\Gamma(L)\Delta Y_t$, $t=p+2,\dots,T$. Then, under $P^{(T)}_{0,0,0;\mu,\Gamma,f}$ and as $T\to\infty$, we have
\begin{align}\label{eqn:rankprocessconvergence_Bg}
\left[ \WTe, \WTf, \BTg \right]^\prime
\wto
\left[ \We, \Wf, \Bg \right]^\prime,
\end{align}
and 
\begin{align} \label{eqn:stochasticintegralconvergence_Bg}
\int_0^1 \We^{(T)}(s-)\rd\Bg^{(T)}(s) \wto \int_0^1 \We(s)\rd\Bg(s). 
\end{align}

Here, $\Bg$ is the associated Brownian bridge of $\Wg$, which itself is a Brownian motion defined on the same probability space $(\Omega,\mathcal{F},\prob_{0,0,0})$ as $\We$ and $\Wf$, with covariance matrix
\begin{align}\label{eqn:covariancematrix}
\cov_{0,0,0} 
\begin{bmatrix} \We(1) \\ \Wf(1) \\ \Wg(1) \end{bmatrix} 
=
\begin{pmatrix} 1 & 1   & \coveg \\ & J_f & J_{fg} \\ & & J_g \end{pmatrix},
\end{align}
where
\begin{align*}
\coveg =&~ \sigma_f^{-1}\sigma_g\int_0^1 F^{-1}(u) \phi_g(G^{-1}(u))\rd u, \\
J_{fg} =&~ \sigma_f\sigma_g\int_0^1 \scoref(F^{-1}(u)) \phi_g(G^{-1}(u))\rd u.
\end{align*}
\item[(ii)] For the case $f\in\Fs$, consider the partial sum process
\begin{align} \label{eqn:partsum_Wg}
\WTg(s) =&~ \frac{1}{\sqrt{T}}\sum_{t=p+2}^{\lfloor sT\rfloor}s_t \sigma_g \left( \phi_{g}\left(G^{-1}\left(\frac{1}{2} + \frac{R_t^{+}}{2(T-p)}\right)\right) \right)
\end{align}
for $s\in[0,1]$, where $s_t$ and $R_t^{+}$ denote the sign of $\Gamma(L)\Delta Y_t$ and the rank of its absolute value, respectively, for $t=p+2,\dots,T$. Then, under $P^{(T)}_{0,0,0;\mu,\Gamma,f}$ and as $T\to\infty$, we have
\begin{align}\label{eqn:rankprocessconvergence_Wg}
\left[\WTe, \WTf, \WTg  \right]^\prime
\wto
\left[ \We, \Wf, \Wg \right]^\prime,
\end{align}
and
\begin{align} \label{eqn:stochasticintegralconvergence_Wg}
\int_0^1 \We^{(T)}(s-)\rd\Bg^{(T)}(s) \wto \int_0^1 \We(s)\rd\Bg(s),
\end{align}
where the law of $\Wg$ is given in (\ref{eqn:covariancematrix}).
\end{itemize}
\end{lemma}

In practice, the autoregressive structure $\Gamma(L)$ will not be known. In that case, one must rely on ranks (and signs) based on innovations calculated using an \emph{estimated} autoregressive structure, i.e., using residuals. This is usually referred to as inference based on \emph{aligned ranks}. For this purpose, we first introduce the following assumption on the estimation of $\Gamma(L)$ (see \citet{HallinPuri1994}). 

\begin{assumption} \label{ass:Gammahat}
\begin{itemize}
\item[(i)] There exists, under the null hypothesis, a $\sqrt{T}$-consistent estimator $(\widehat\Gamma_{1},\dots,\widehat\Gamma_{p})^\prime$ of $(\Gamma_{1},\dots,\Gamma_{p})^\prime$. That is, for all $\mu\in\SR$, $\Gamma\in\Gspace$, $f\in\F$ and all $\epsilon>0$, there exists $b$ and $T_b$ such that 
\begin{align}P^{(T)}_{0,0,0;\mu,\Gamma,f}\left\{\|\sqrt{T}(\widehat\Gamma - \Gamma)\|>b\right\} < \epsilon, ~~~ \forall t\geq T_b.
\end{align}
\item[(ii)] The estimator $(\widehat\Gamma_{1},\dots,\widehat\Gamma_{p})^\prime$ is discretized on grids of mesh width $T^{-1/2}$. That is, the number of possible values of $(\widehat\Gamma_{1},\dots,\widehat\Gamma_{p})^\prime$ in balls of the form $\{\Upsilon\in\SR^p ~ \| ~ T^{-1/2}(\Upsilon-\Upsilon_0)\leq c\}$ remains bounded, as $T\to\infty$, for all $\Upsilon_0\in\SR^p$ and all $c>0$.
\end{itemize}
\end{assumption}
Part~(i) of Assumption~\ref{ass:Gammahat} is mild as many known estimators of  $(\Gamma_{1},\dots,\Gamma_{p})^\prime$ exist for the AR($p$) model that can be applied to the increments $\Delta Y_t$, see, e.g., \citet{BrockwellDavis2002}. Part~(ii) is standard in the semiparametric literature and can easily be met by transforming the estimator in Part~(i), see, e.g., \citet{Bickel1982} and \citet{Kreiss1987}. Such an estimator is often called \textit{locally asymptotically discrete}.
The use of aligned ranks does not invalidate the conclusion of Lemma~\ref{lem:rankprocess}. This is the content of the following result, of which the proof is provided in \ref{supplementA}.

\begin{lemma} \label{lem:rankprocess_aligned}
Under Assumption~\ref{ass:Gammahat}, Lemma~\ref{lem:rankprocess} remains valid in case aligned signs and ranks are used, i.e., signs and ranks of innovations calculated using the estimated autoregressive coefficients $(\widehat\Gamma_{1},\dots,\widehat\Gamma_{p})$.
\end{lemma}

We denote by $\BTga$ and $\WTga$ the aligned-rank-based counterparts of the rank-based processes $\BTg$ and $\WTg$, respectively. 

\subsection{Hybrid rank tests based on a reference density}\label{sec:hybrid_rank_tests}

\noindent In this section, we propose our unit root tests, for both the non-symmetric case ($f\in\F$) and the symmetric case ($f\in\Fs$). Section~\ref{sec:invariance} provides the basis for optimal invariant tests in the limit experiment $\limitE(f)$ and $\limitEs(f)$. Lemmas~\ref{lem:rankprocess} and~\ref{lem:rankprocess_aligned} can subsequently be used to approximate, in the sequences of unit root experiments $\limitET(f)$ and $\limitETs(f)$, the observable processes in these limit experiments. More specifically, these two lemmas indicate that constructing the rank-based score partial sum processes $\BTg$ and $\WTg$ as in (\ref{eqn:partsum_Bg}) and (\ref{eqn:partsum_Wg}), together with $\WTe$, intuitively corresponds to observing the $\sigma$-fields $\Invg := \sigma(\We, \Bg)$ and $\Invgs := \sigma(\We, \Wg)$ in the limit experiments $\limitE(f)$ and $\limitEs(f)$, respectively. This then leads to our asymptotically invariant tests based on $\Invg$ and $\Invgs$.

The following proposition establishes the likelihood ratio restricted to the information $\Invg$ and $\Invgs$. The Neyman-Pearson lemma implies that tests based on these likelihood ratios are point optimal amongst the class of invariant tests in the limit experiments.
\begin{proposition} \label{prop:likelihoodratios} 
Define $\Wind$ implicitly via the decomposition 
\begin{equation}\label{eqn:decomposition}
\Wg = \coveg\We + \sqrt{J_g - \coveg^2}\Wind,
\end{equation}
which is a standard Brownian motion under $\mathbb{P}_{0,0,0}$ and denote the associated bridge by $\Bind$.\footnote{The implicit requirement $J_g \geq \coveg^2$ in the decomposition (\ref{eqn:decomposition}) is directly guaranteed the Cauchy-Schwarz inequality, $\sigma_g^2\int_0^1\left|\score_g(G^{-1}(u))\right|^2\rd u \cdot \sigma_f^{-2}\int_0^1 \left|F^{-1}(u)\right|^2 \rd u \geq \big|\sigma_f^{-1}\sigma_g\int_0^1 F^{-1}(u) \phi_g(G^{-1}(u))\rd u\big|^2$, and the fact that $\sigma_f^{-2}\int_0^1 [F^{-1}(u)]^2 \rd u = 1$.}
\begin{itemize}
\item[(i)]
The likelihood ratio $\rd\mathbb{P}_h^\maxInvf/\rd\mathbb{P}_0^\maxInvf$ restricted to the outcome space $\Invg$ is given by
\begin{align}\label{eqn:loglikelihoodratioInvg}
\frac{\rd\mathbb{P}^{\Invg}_h}{\rd\mathbb{P}^{\Invg}_0}
 = \mathbb{E}_0\left[\frac{\rd \prob_{h}^\maxInvf } {\rd \prob_{0}^\maxInvf}\mid\Invg\right]=\exp\left(h\EffCSg -\frac{1}{2}h^2\EffFIg\right),
\end{align}
where
\begin{align*}
\EffCSg =&~ \EffCSe + \lambda\EffCSind, \\
\EffFIg =&~ \int_0^1\We^2(s)\rd s + \lambda^2\left(\frac{J_g}{\coveg^2}-1\right)\left[\int_0^1\We(s)^2\rd s - \left( \int_0^1 \We(s) \rd s\right)^2\right],
\end{align*}
with $\EffCSe = \int_0^1\We(s)\rd\We(s)$, $\EffCSind = \sqrt{J_g/\coveg^2-1}\int_0^1\We(s)\rd\Bind(s)$, and 
\begin{align*}
\lambda=(J_{fg}\coveg-\coveg^2)/(J_g-\coveg^2).
\end{align*}
\item[(ii)]
The likelihood ratio $\rd\mathbb{P}^{\Invgs}_h/\rd\mathbb{P}^{\Invgs}_0$ restricted to the outcome space $\Invgs$ is given by
\begin{align}\label{eqn:loglikelihoodratioInvgs}
\frac{\rd\mathbb{P}^{\Invgs}_h}{\rd\mathbb{P}^{\Invgs}_0}
 =&~ \mathbb{E}_0\left[\frac{\rd \prob_{h}^{\maxInvfs} } {\rd \prob_{0}^{\maxInvfs}}\mid\Invg\right]=\exp\left(h\EffCSgs -\frac{1}{2}h^2\EffFIgs\right),
\end{align}
where
\begin{align*}
\EffCSgs =&~ \EffCSe + \lambda\EffCSinds, \\
\EffFIgs =&~ \left(1 + \lambda^2\frac{J_g}{\coveg^2}-\lambda^2\right)\left[\int_0^1\We(s)^2\rd s\right],
\end{align*}
with $\EffCSinds = \sqrt{J_g/\coveg^2-1}\int_0^1\We(s)\rd\Wind(s)$.
\end{itemize}
\end{proposition}

\begin{remark} To construct the test statistic in the limit experiment $\limitE(f)$, instead of simply replacing $\Bf$ by $\Bg$, we rely on the likelihood ratio of $\Invg$. This, by the Neyman-Pearson Lemma, is the optimal way regardless of complexity. Clearly, it has $\Invg\subseteq\maxInvf$ so that $\Invg$ is invariant for the group of transformations $g_\eta$.\footnote{This is due to the decomposition $\Bg=\coveg\Be+\sum_{k=1}^{\infty}J_{g,k}\Bbk$.} When $g=f$, $\Invg=\maxInvf$ so that it is maximally invariant, which means that we capture all available information about $h$, and this in turn gives the asymptotic optimality for the finite-sample counterpart below. The same argument also holds for the limit experiment $\limitEs(f)$.
\end{remark}

\begin{remark}\label{remark:limitexperimentwithreferencedensity}
The result of Proposition~\ref{prop:likelihoodratios} can also be achieved by first applying Girsanov's Theorem to the experiment associated to observing $\We$ and $\Wg$ defined via
\begin{align*}
\rd \We(s) &= h \We(s)\rd s + \rd Z_\varepsilon(s), \\ 
\rd \Wg(s) &= h  J_{fg}\We(s)\rd s + \rd \Zg(s), 
\end{align*}
to get the likelihood ratio of $\Invgs$. This experiment is obtained by combining the limit experiment in Proposition~\ref{prop:Girsanov} and the covariance matrix in (\ref{eqn:covariancematrix}). In other words, this provides the limit experiment associated to $\Invgs$. Subsequently, one can take the expectation of the likelihood ratio of $\Invgs$ obtained above conditional on $\Invg$ to get the likelihood ratio of $\Invg$. The associated limit experiment is given by
\begin{align*}
\rd \We(s) &= h \We(s)\rd s + \rd Z_\varepsilon(s), \\ 
\rd \Bg(s) &= h  J_{fg}\left[\We(s) - \overline{\We}\right]\rd s + \rd \left[\Zg(s) - s \Zg(1)\right],
\end{align*}
where $\overline{\We} = \int_0^1\We(r)\rd r$. 
\end{remark}

Observe that $\Wind$ is a standard Brownian motion under $\mathbb{P}_{0,0,0}$ and independent of $\We$. When $g=f$, we have $J_{fg}=J_f=J_g$ and $\coveg=1$, so that $\lambda=1$ and $\Bg=\Bf$. As a result, we have $\EffCSg=\EffCS_f$ and $\EffFIg=\EffFI_f$, and $\EffCSgs=\CSlim_f$ and $\EffFIgs=\FIlim_f$. 

The central idea to construct a hybrid rank test is to use a (quasi)-likelihood ratio test based on $\ALR_{\Invg}(h,\lambda) := h\EffCSg -\frac{1}{2}h^2\EffFIg$ from (\ref{eqn:loglikelihoodratioInvg}) for the case of $f\in\F$, and $\ALR_{\Invgs}(h,\lambda) := h\EffCSgs -\frac{1}{2}h^2\EffFIgs$ from (\ref{eqn:loglikelihoodratioInvgs}) for the case of $f\in\Fs$. In both cases, we then replace $\We$ and $\Bg$ by their finite-sample counterparts in Lemma~\ref{lem:rankprocess_aligned} (or those in Lemma~\ref{lem:rankprocess} in case $\Gamma(L)$ would be known, for example, when $p=0$). The remaining unknown finite-sample parameters $\sigma_f^2$ and $\lambda$ are replaced by estimates that need to satisfy the following condition.
\begin{assumption}\label{ass:estimator_variance}
There exist consistent, under the null hypothesis, estimators $\hat\sigma_f^2>0~a.s.$, $\statcoveg$ and $\statJfg$ of $\sigma_f^2$, $\coveg$ and $J_{fg}$, respectively. More precisely, for all $\mu\in\SR$, $\Gamma\in\Gspace$, $f\in\F$, we have $\hat\sigma_f^2 \pto \sigma_f^2$, $\statcoveg \pto \coveg$, and $\statJfg \pto J_{fg}$, under $\law{0,0,0;\mu,\Gamma,f}$ as $T\to\infty$. 
\end{assumption}

Such estimators are easily constructed, although $\hat{J}_{fg}$ is somewhat more involved. Estimating the \emph{real-valued} cross-information $J_{fg}$ requires nonparametric techniques, but is considerably simpler than a full \emph{nonparametric} estimation of $\scoref$. Estimating $J_{fg}$ can be done along similar lines as estimating the Fisher information $J_f$, see, e.g., \citet{Bickel1982}, \citet{BKRW}, \citet{Schick1986}, and \citet{Klaassen1987}. A direct rank-based estimator of $J_{fg}$ has been proposed in \citet{CassartHallinPain2010}. It is also worth noting that the consistency automatically also holds under local alternatives due to Le Cam's third lemma.


Based on a chosen reference density $g$ satisfying Assumption~\ref{ass:ref_density} and estimators $\hat\sigma_f$, $\statcoveg$ and $\statJfg$ satisfying Assumption~\ref{ass:estimator_variance}, we introduce the following partial sum processes: 
\begin{align*}
\statWTe(s)   =&~ \frac{1}{\sqrt{T}}\sum_{t=p+2}^{\lfloor sT\rfloor} \frac{\widehat\Gamma(L)\Delta Y_t}{\hat\sigma_f}, \\
\statBTind(s) =&~ \left(\frac{J_g}{\statcoveg^2}-1\right)^{-\frac{1}{2}}\left[\frac{\BTga(s)}{\statcoveg}-\left(\statWTe(s) - s \statWTe(1) \right)\right], \\
\statWTind(s) =&~ \left(\frac{J_g}{\statcoveg^2}-1\right)^{-\frac{1}{2}}\left[\frac{\WTga(s)}{\statcoveg}-\statWTe(s)\right]
\end{align*}
where $\BTga(s)$ and $\WTga(s)$ are defined by Lemma~\ref{lem:rankprocess_aligned}. Note that in the case of known $\Gamma(L)$, e.g., the i.i.d.\ case, one can simply use $\widehat{\Gamma}(L) = \Gamma(L)$ (where $\BTga(s)=\BTg(s)$ and $\WTga(s)=\WTg(s)$). Now, given a fixed alternative $\bar{h}<0$, we define
\begin{align*}
\statLRmg(\bar{h},\hat\lambda) := \bar{h}\statCS-\frac{1}{2}\bar{h}^2\statFI,
\end{align*}
with
\begin{align*}
\statCS =&~ \statDe + \hat\lambda\statDind, \\
\statFI =&~ \int_0^1\left(\statWTe(s-)\right)^2\rd s + \hat{\lambda}^2\left(\frac{J_g}{\statcoveg^2}-1\right) \\
&~ \times \left[\int_0^1\left(\statWTe(s-)\right)^2\rd s-\left(\int_0^1\statWTe(s-)\rd s\right)^2\right],
\end{align*}
where $\statDe = \int_0^1\statWTe(s-)\rd\statWTe(s)$, $\statDind = \sqrt{J_g/\statcoveg^2-1}\int_0^1\statWTe(s-)\rd\statBTind(s)$ and $\hat\lambda=(\statJfg\statcoveg-\statcoveg^2)/(J_g-\statcoveg^2)$. Define also for the case of $f\in\Fs$, 
\begin{align*}
\statLRmgs(\bar{h},\hat\lambda) := \bar{h}\statCSs-\frac{1}{2}\bar{h}^2\statFIs,
\end{align*}
with
\begin{align*}
\statCSs =&~ \statDe + \hat\lambda\statDinds, \\
\statFIs =&~ \bigg(1 + \hat{\lambda}^2\frac{J_g}{\statcoveg^2} - \hat{\lambda}^2\bigg)\left[\int_0^1\left(\statWTe(s-)\right)^2\rd s\right],
\end{align*}
where $\statDinds = \sqrt{J_g/\statcoveg^2-1}\int_0^1\statWTe(s-)\rd\statWTind(s)$. 

By Slutsky's theorem, we have the convergences $\big(\statWTe, \statBTind \big)^\prime \wto \big( \We, \Bind \big)^\prime$ and $\big(\statWTe, \statWTind \big)^\prime \wto \big( \We, \Wind \big)^\prime$, and the convergences of stochastic integrals $\int_0^1\statWTe(s-)\rd\statBTind(s) \wto \int_0^1\We(s)\rd\Bind(s)$ and $\int_0^1\statWTe(s-)\rd\statWTind(s) \wto \int_0^1\We(s)\rd\Wind(s)$ (see the proof of Lemma~\ref{lem:rankprocess}). We thus also obtain $\statLRmg(\bar{h},\hat\lambda)\wto\ALR_{\Invg}(\bar{h},\lambda)$ and $\statLRmgs(\bar{h},\hat\lambda)\wto\ALR_{\Invgs}(\bar{h},\lambda)$ under $P^{(T)}_{0,0,0;\mu,\Gamma,f}$. Define the critical values $\cvmg(\bar{h},\coveg,\lambda,J_g;\alpha)$ and $\cvmgs(\bar{h},\coveg,\lambda,J_g;\alpha)$ by the $(1-\alpha)$-quantiles of $\ALR_{\Invg}(\bar{h},\lambda)$ and $\ALR_{\Invgs}(\bar{h},\lambda)$, respectively. This leads to the (feasible) tests
\begin{align*}
\Ftf_{\Invg}(\bar{h},\alpha):=\indicator{\statLRmg(\bar{h},\hat\lambda)\geq \cvmg(\bar{h},\statcoveg,\hat\lambda,J_g;\alpha)},
\end{align*}
and 
\begin{align*}
\Ftf_{\Invgs}(\bar{h},\alpha):=\indicator{\statLRmgs(\bar{h},\hat\lambda)\geq \cvmgs(\bar{h},\statcoveg,\hat\lambda,J_g;\alpha)}.
\end{align*}
These tests are not only based on the ranks of $\Delta Y_t$ but also their average, therefore, we name them Hybrid Rank Tests (HRTs). 

We can now state our main theoretical result.
\begin{theorem} \label{thm:propertiesHRT}
\begin{itemize}
\item[(i)] Under the Assumptions 1 - 5, for each chosen $\alpha\in (0,1)$ and $\bar{h}\in(-\infty,0)$, we have
\begin{itemize}
\item[(1)] The HRT $\Ftf_{\Invg}(\bar{h},\alpha)$ is asymptotically of size $\alpha$.
\item[(2)] The HRT $\Ftf_{\Invg}(\bar{h},\alpha)$ is asymptotically invariant.	
\item[(3)] The HRT $\Ftf_{\Invg}(\bar{h},\alpha)$ is point-optimal, at $h=\bar{h}$, if $g=f$.
\end{itemize}
\item[(ii)]
Under the Assumptions 1 - 5 and $g\in\Fs$, for each chosen $\alpha\in (0,1)$ and $\bar{h}\in(-\infty,0)$, we have
\begin{itemize}
\item[(1)] The HRT $\Ftf_{\Invgs}(\bar{h},\alpha)$ is asymptotically of size $\alpha$.
\item[(2)] The HRT $\Ftf_{\Invgs}(\bar{h},\alpha)$ is asymptotically invariant.	
\item[(3)] The HRT $\Ftf_{\Invgs}(\bar{h},\alpha)$ is point-optimal, at $h=\bar{h}$, if $g=f$.
\end{itemize}
\end{itemize}
\end{theorem}

Theorem~\ref{thm:propertiesHRT} shows the HRTs are valid irrespective of the choice of the reference density and point-optimal for a correctly specified reference density. Moreover, in the corollary below, we state that the HRTs enjoy a Chernoff-Savage type result.

\begin{corollary}\label{cor:chernoff-savage}
Fix $\alpha\in(0,1)$ and $\bar{h}<0$. The HRT $\Ftf_{\Invg}(\bar{h},\alpha)$ is, for any reference density $g$ satisfying Assumption~\ref{ass:ref_density}, more powerful, at $h=\bar{h}$ and for $\mu\in\SR$, $\Gamma\in\Gspace$ and $f\in\F$, than the ERS test except when $f$ is Gaussian where they have equal powers. The same argument holds for the HRT $\Ftf_{\Invgs}(\bar{h},\alpha)$ with $g\in\Fs$ for any $f\in\Fs$. 
\end{corollary}

Corollary~\ref{cor:chernoff-savage} is a particularly useful result for applied work. The HRT dominates its classical canonical Gaussian counterpart, i.e., the ERS test in the present model, for any reference density $g$. Traditionally, this claim can only be made for Gaussian reference densities, but the framework here even allows for a stronger result. Our formulation of the testing problem using invariance arguments is convenient in this respect: the larger the invariant $\sigma$-field that is used, the more powerful the test.

The situation can be compared to Quasi Maximum Likelihood methods. However, again, in classical situations these methods are restricted to Gaussian reference densities. In the present setup, any reference density $g$ (subject to the regularity conditions imposed) can be used. The resulting test will always be valid,  but more powerful in case the reference density chosen is closer to the true underlying density $f$.

\begin{remark} \label{remark:maximalinvariant}
It is worth noting that the invariance constraint is only imposed in the limit and, therefore the maximal invariant needs only to be derived in the limit experiment. In other words, in the finite-sample unit root testing experiment $\limitET(f)$ (or $\limitETs(f)$ for $f\in\Fs$), we actually use statistics that are only asymptotically invariant (i.e., their limiting equivalents are measurable with respect to the maximally invariant sigma-field $\maxInvf$ (or $\maxInvfs$ for $f\in\Fs$), while, for finite $T$, they are not necessarily (maximally) invariant with respect to some transformation on the density $f$. In fact, a maximal invariant may very well not even exist in the finite-sample experiments. Specifically, $\WTe$ approximates $\We$ in the limit, whose distribution does not change with the density $f$, while the distribution of $\We^{(T)}$ does depend on the density $f$. Instead, the statistic $\BTg$ (or $\WTg$ for $f\in\Fs$) is distribution-free, that is, its distribution is not affected by any transformation on the density $f$. In Section~\ref{sec:MonteCarlo} below, we will show that the asymptotic approximations work well even in smaller samples.
\end{remark}

\begin{remark} \label{remark:robustefficiencytradeoff}
The additional power of the HRT compared to the ERS test is not free due to the stronger weak convergence assumption employed. Consequently, the class of models for which the HRTs are valid forms a sub-class of the class where the ERS tests are valid. In this sub-class, the HRT dominates the ERS test, but outside they may even loose validity. In the opposite direction, the \citet{MullerWatson2008} low-frequency unit root test can be applied in a even larger class of models than the ERS tests. Again, within the class of models where the ERS test is valid, it has lower power. A more general and detailed discussion in this direction can be found in \citet{Muller2011}. Our test will still be relevant in many applications, notably those where policy implications are derived under an i.i.d.\ assumption on the innovations. Also, our approach can most likely be extended to situations where the innovations are described by some explicit dynamic location-scale model. We come back to this point in Section~\ref{sec:conclusion}.

\end{remark}

\subsection{Approximate hybrid rank tests}\label{sec:approximatehybridranktests}

\noindent A somewhat inconvenient aspect of the hybrid rank tests is that we need to estimate ``the real-valued parameter'' $J_{fg}$. As mentioned before, this is (much) less complicated than estimating the score function $\scoref$ (as needed in \citet{Jansson2008}), but might still be considered cumbersome, despite all references mentioned below Assumption~\ref{ass:estimator_variance}. Moreover, the critical value $\cvmg(\bar{h},\statcoveg,\hat\lambda,J_g;\alpha)$ depends on the estimates $\statcoveg$ and $\hat\lambda$ (henceforth $\statJfg$). Of course, this introduces no difficulty to implementing the test for a single dataset (though one would need to simulate a critical value), however, when it comes to a Monte Carlo study to access the performances of the HRTs, the computational effort will be significant. Therefore, we introduce a simplified version of the hybrid rank test. This simplified test is obtain by invoking $\lambda=1$, which holds in case $g=f$. 

To be precise, define
\begin{align} \label{eqn:teststatistics_AHRT}
\statLRg(\bar{h}) := \statLRmg(\bar{h},1) = \bar{h}\statCS - \frac{1}{2}\bar{h}^2\statFI,
\end{align}
where
\begin{align*}
\statCS=&~
\frac{1}{\statcoveg}\int_0^1\statWTe(s-)\rd\statBTg(s) + \statWTe(1) \int_0^1 \statWTe(s-)\rd s, \\
\statFI=&~
\frac{J_g}{\statcoveg^2}\int_0^1\left(\statWTe(s-)\right)^2\rd s - \left(\int_0^1\statWTe(s-)\rd s\right)^2 \left(\frac{J_g}{\statcoveg^2}-1\right),
\end{align*}
and \begin{equation}
\statLRgs(\bar{h}) := \statLRmgs(\bar{h},1) = \bar{h}\statCSs - \frac{1}{2}\bar{h}^2\statFIs,
\end{equation}
where
\begin{align*}
\statCSs=&~
\frac{1}{\statcoveg}\int_0^1\statWTe(s-)\rd\WTga(s), \\
\statFIs=&~
\frac{J_g}{\statcoveg^2}\int_0^1\left(\statWTe(s-)\right)^2\rd s.
\end{align*}
Also define $L_g(\bar{h}):=\ALR_{\Invg}(\bar{h},1)$ and $L_g^{\mathbb{S}}(\bar{h}):=\ALR_{\Invgs}(\bar{h},1)$, then we have $\statLRg(\bar{h})\wto L_g(\bar{h})$ and $\statLRgs(\bar{h})\wto L_g^\mathbb{S}(\bar{h})$ under $\law{0,0,0;\mu,\Gamma,f}$. Denoting the $(1-\alpha)$-quantiles of $L_g(\bar{h})$ and $L_g^\mathbb{S}(\bar{h})$ by $\cvg(\bar{h},\coveg,J_g;\alpha)$ and $\cvgs(\bar{h},\coveg,J_g;\alpha)$, respectively. These lead to the feasible tests
\begin{align*}
\Ftf_g(\bar{h},\alpha):=\indicator{L_g(\bar{h})\geq \cvg(\bar{h},\statcoveg,J_g;\alpha)},
\end{align*}
and
\begin{align*}
\Ftf_g^\mathbb{S}(\bar{h},\alpha):=\indicator{L_g^\mathbb{S}(\bar{h})\geq \cvgs(\bar{h},\statcoveg,J_g;\alpha)}.
\end{align*}
Since $\Ftf_g(\bar{h},\alpha)$ and $\Ftf_g^\mathbb{S}(\bar{h},\alpha)$ are approximate versions of the Hybrid Rank Tests $\Ftf_{\Invg}(\bar{h},\alpha)$ and $\Ftf_{\Invgs}(\bar{h},\alpha)$, we refer to them as Approximate Hybrid Rank Tests (AHRTs).

\begin{theorem} \label{thm:propertiesAHRT}
Under the same conditions as Theorem~\ref{thm:propertiesHRT}, the asymptotic properties of the Hybrid Rank Tests --- validity, invariance, and point-optimality when $g=f$ --- also hold for the Approximate Hybrid Rank Tests.
\end{theorem}
The proof of Theorem~\ref{thm:propertiesAHRT} follows along the same lines as that of Theorem~\ref{thm:propertiesHRT} but using the weak convergences $\statLRg(\bar{h}) \wto L_g(\bar{h})$ and $\statLRgs(\bar{h}) \wto L_g^\mathbb{S}(\bar{h})$. The simulation results in Section~\ref{sec:MonteCarlo} show that these asymptotic properties carry over to finite samples.

\begin{remark} \label{remark:ChernofSavage}
Although we are not able to provide a rigorous mathematical proof, the Monte-Carlo study indicates that the Chernoff-Savage property is also preserved for the AHRTs, at least in case the reference density $g$ is chosen to be Gaussian. Such a result would be more in line with applications of the Chernoff-Savage result in classical LAN situations.
\end{remark}

From a computational point of view, the AHRTs have the advantage that nonparametric estimation of $J_{fg}$ is no longer needed. This significantly reduces the computational effort in the Monte Carlo study. Indeed, even though the critical value $\cvg(\bar{h},\coveg,J_g;\alpha)$ and $\cvgs(\bar{h},\coveg,J_g;\alpha)$ are still data dependent, it is, for given $\alpha$, $\bar{h}$, and reference density $g$, a function of only one argument --- the parameter $\coveg$. Observe, by Cauchy-Schwarz, that $\coveg$ is bounded by $\sqrt{J_g}$. For the chosen three reference densities, the critical value functions are listed in Table~\ref{table:criticalvalues}. These are obtained by fitting a fourth-order polynomial to the exact critical values. In the Monte Carlo study (Section~\ref{sec:MonteCarlo}), we use these approximating critical value functions for computational speed.

\begin{table}[!h]
\caption{This table provides estimated critical value functions for three reference densities: Gaussian ($J_g=1$), Laplace ($J_g=2$), and Student $t_3$ ($J_g=2$) at $\alpha=5\%$ and $\bar{h}=-7\coveg$. For each case, the critical value function is estimated by OLS using simulated critical values on the interval $[0,\sqrt{J_g}]$ with a grid where adjacent points are 0.01 apart.}\label{table:criticalvalues}
\begin{tabular}{c|c}
\hline\hline
{\it g}    & \\ \hline
Gaussian   & $\cvg(-7\coveg,\coveg,1; 5\%)$ = $0.96+1.88\coveg-3.98\coveg^2+6.74\coveg^3-5.45\coveg^4$            \\
\hline
Laplace    & $\cvg(-7\coveg,\coveg,2; 5\%)$ = $0.25+2.30\coveg-3.58\coveg^2+4.30\coveg^3-2.45\coveg^4$            \\
\hline
Student $t_3$    & $\cvg(-7\coveg,\coveg,2; 5\%)$ = $0.25+2.30\coveg-3.58\coveg^2+4.30\coveg^3-2.45\coveg^4$            \\
\hline
\end{tabular}
\end{table}

\begin{remark}[Nonparametrically estimated reference density] \label{remark:extension_nonparametric}
The Hybrid Rank Test and the Approximate Hybrid Rank Test are optimal when the reference density $g$ coincides with the actual innovation density $f$. It is therefore reasonable to consider these test using a nonparametric estimate of $f$, say $\hat{f}$, as reference density. Commonly such estimators are based on the order statistics of the residuals $\hat\varepsilon_t$. Under a suitable consistency condition, the HRT based on $\hat{f}$ asymptotically is conjectured to behave as the HRT based on the true innovation density $f$. Thus, such test achieves the optimality properties of Theorem \ref{thm:propertiesHRT} and Theorem \ref{thm:propertiesAHRT} globally. Notably, even if there exists relatively large bias in the estimation of $f$, the usage of rank statistics ensures zero expectation of the feasible score function $\phi_{\hat{f}}[\widehat{F}^{-1}(R_t/(T+1))]$, which furthermore ensures the validity of the HRTs and the AHRTs. 


\end{remark}

\section{Monte Carlo study}\label{sec:MonteCarlo}
\noindent
This section reports the results of a Monte Carlo study to corroborate our asymptotic results, and to analyze the small-sample performance of the Approximate Hybrid Rank Tests. As mentioned earlier, we use the Approximate Hybrid Rank Tests in this simulation to avoid having to simulate the critical value for each individual replication. For the fixed alternative, we choose $\bar{h}=-7\coveg$ for two reasons. First, when $g=f$, we have $\coveg=1$ and hence $\bar{h}=-7$, which is in line with \citet{ERS1996}. 
Second, as $\statcoveg$ appears in the denominator in the AHRT statistic (\ref{eqn:teststatistics_AHRT}), the statistic becomes better behaved for values of $\statcoveg$ close to zero.
The corresponding critical value functions for various reference densities are provided in Table~\ref{table:criticalvalues}. The estimators for $\sigma_f^2$  and $\coveg$ we use are 
\begin{align*} 
\hat\sigma_f^2
 =&~
\frac{1}{T-p-1}\sum_{t=p+2}^T \left(\widehat\Gamma(L)\left(\Delta Y_t - \frac{1}{T}\sum_{t=p+2}^T\Delta Y_t\right)\right)^2,\\
\statcoveg
 =&~
\frac{1}{T-p-1}\sum_{t=p+2}^T \frac{\widehat\Gamma(L)\Delta Y_t}{\hat{\sigma}_f}
	\sigma_g\phi_g\left(G^{-1}\left(\frac{\hat{R}_t}{T-p} \right)\right).
\end{align*}
Moreover, to simplify the notation, we denote the Approximate Hybrid Rank Test with reference density $g$ by ${\rm AHRT}$-$g$ and, in particular, by ${\rm AHRT}$-$\phi$ for Gaussian reference density and by ${\rm AHRT}$-$\hat{f}$ for a (nonparametrically) estimated reference density. To be specific, for the nonparametric estimation of $f$, we employ a kernel-based method $\hat{f}(x) = (T\mathfrak{h})^{-1}\sum_{t=1}^{T}K\left((x-\hat{\ee}_t)/\mathfrak{h}\right)$, where the kernel $K$ is chosen to be Gaussian, $\mathfrak{h}$ is the bandwidth chosen by the rule $\left(4/(3T)\right)^{\frac{1}{5}}\hat\sigma_f$, and $\hat{\ee}_t$ are the residuals from the regression (\ref{eqn:arma_regression}) below. Throughout we use significance level $\alpha = 5\%$ and all results are based on 20,000 Monte-Carlo replications.

We compare the performances of the propsed AHRT tests with two alternatives. First we consider the Dickey-Fuller test (denoted by DF-$\rho$) from \citet{DickeyFuller79}. This test is based on the statistic $T(\hat{\rho}-1)$ where $\hat{\rho}$ is the least-squares estimator in the regression 
\begin{align} \label{eqn:arma_regression}
Y_t=\mu+\rho Y_{t-1}+ \sum_{i=1}^{p}\Gamma_i\Delta Y_{t-i} + \varepsilon_t. 
\end{align}
The critical values for this test are -13.52 for $T=100$ and -14.05 for $T=2,500$. The second competitor is the ERS test with $\bar{h}=-7$. This test is based on the statistic $[S(\bar{\alpha})-\bar{\alpha}S(1)]/\hat{\omega}^2$ with $\bar{\alpha}=1+T^{-1}\bar{h}$ and $S(a) = (Y_a-Z_a\hat\beta)'(Y_a-Z_a\hat\beta)$, with $Y_a$ and $Z_a$ defined as
\begin{align*}
Y_a =&~ (Y_1,Y_2-aY_1,\dots,Y_T-aY_{T-1})',\\
Z_a =&~ (1,1-a,\dots,1-a)',
\end{align*}
where $\hat\beta$ is estimated by regressing $Y_{\bar{\alpha}}$ on $Z_{\bar{\alpha}}$. The long-run variance estimator (for ERS test), $\hat{\omega}^2$, is chosen to be $\hat{\omega}^2_{AR(p)} = \hat\sigma_e^2 \big/ \left(1-\sum_{i=1}^p \widehat{\Gamma}_i \right)^2$ with $\hat\sigma_e^2 = \sum_{t=1}^T\hat{\ee}_t^2/T$, where the residuals $\hat{\ee}_t$ and coefficient estimates $\widehat{\Gamma}_i$ are from the regression in (\ref{eqn:arma_regression}). The critical values for this test are 3.11 for $T=100$ and 3.26 for $T=2,500$. We do not consider the Dickey-Fuller $t$-test as it is dominated by the DF-$\rho$ test in the current model. Similarly, the DF-GLS test proposed in \citet{ERS1996} is also omitted as it behaves asymptotically the same as the ERS test, but can be oversized in smaller samples.

\subsection{Simulation results with i.i.d.\ innocations} \label{sec:Montecarlo_iid}
In this section, we start the Monte Carlo study with i.i.d.\ innovations, i.e., the data is generated by the model in (\ref{eqn:DGP1})-(\ref{eqn:DGP3}) with $\Gamma(L) = 1$. Therefore, for the newly-proposed AHRT test and its competitors introduced above, we have $\widehat{\Gamma}(L) = 1$ (or, equivalently, $\widehat{\Gamma}_1 = \cdots = \widehat{\Gamma}_p = 0$).

\subsubsection*{Large-sample performance}\label{sec:AsympPerform}
We first use the large-sample performances to illustrate the asymptotic properties. In particular, the chosen sample size $T$ is $2,500$. 

\begin{figure}[!htb] 
\hspace*{-33mm}
\includegraphics[width=7.5in]{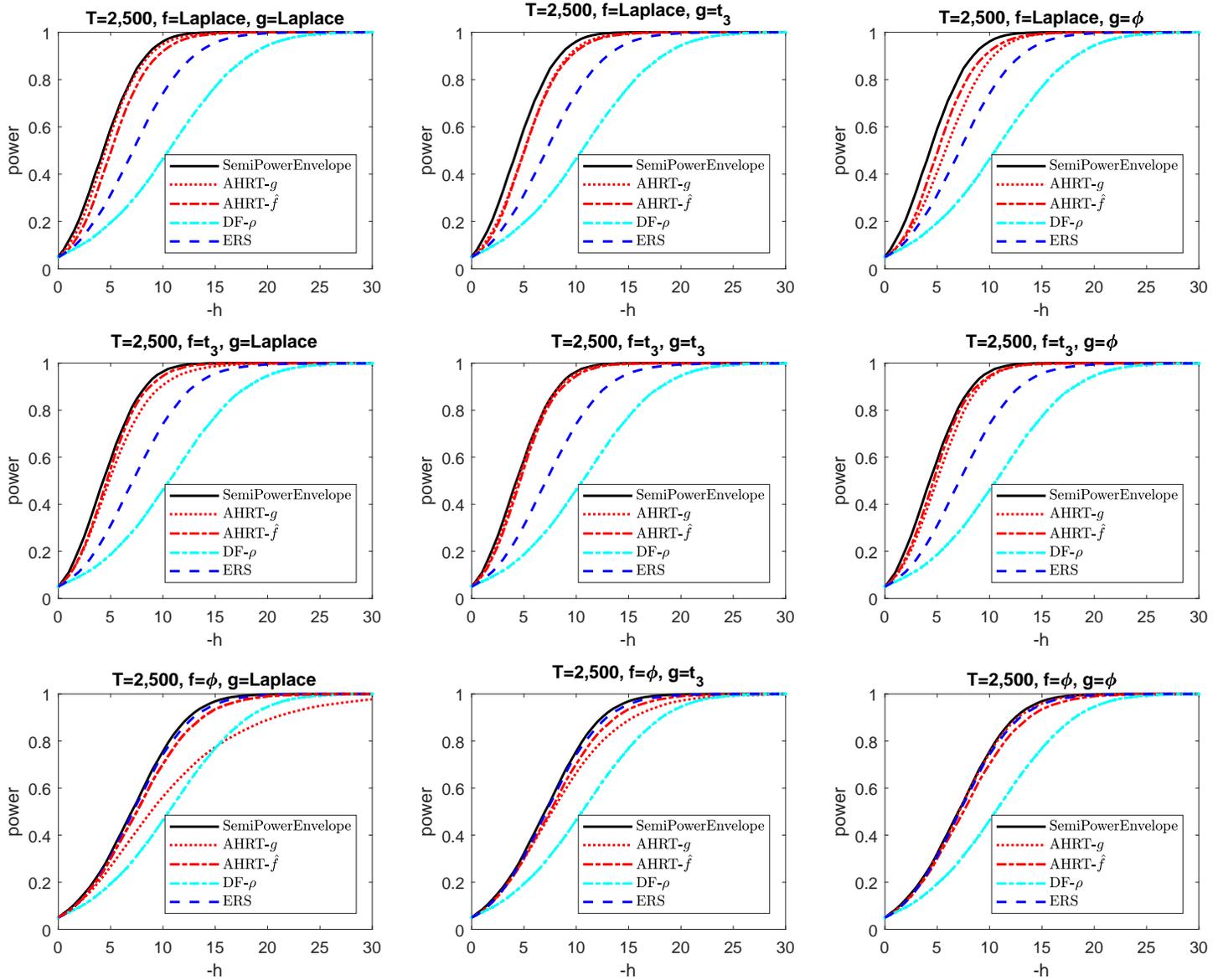}
\caption{IID errors case: large-sample power functions of the {\rm AHRT}-g with various reference densities $g$, the {\rm AHRT}-$\hat{f}$, and other selected unit root tests under the true innovation densities $f$: Gaussian, Laplace, Student's $t_3$.}
\label{figure_n2500}
\end{figure}

Figure~\ref{figure_n2500} shows the power curves for 9 combinations of 3 innovation densities $f$ and 3 reference densities $g$ (for AHRT-$g$): $f$ and $g$ are chosen to be Laplace, Student $t_3$, or Gaussian.\footnote{The semiparametric power envelopes are based on $40,000$ Monte Carlo replications where the W-processes are approximated by a simple Euler approximation using $2,500$ grid points.} In line with our theoretical results, we find that the AHRT-$g$ test outperforms the two competitors in most cases. More specifically, when $g=f$ (the graphs on the diagonal), the AHRT-$f$ has power very close to the semiparametric power envelope and it is tangent to it at the point $-h=7$. This verifies the point-optimal result of the AHRT-$f$ test in Theorem~\ref{thm:propertiesAHRT}. The AHRT-$\hat{f}$ test has a similar behavior as the AHRT-$f$ but with a slightly lower power due to the efficiency loss in estimating the density $f$. This small amount of power loss is also different from case to case, e.g., for $f = t_3$, this power loss is almost indistinguishable; and the amount decreases to zero as $T$ goes to infinity. Moreover, when the reference density $g$ is Gaussian (the three right-most graphs), the AHRT-$\phi$ outperforms its competitors for non-Gaussian $f$; while for Gaussian $f$, the AHRT-$\phi$ test and the ERS test have indistinguishable power (and they outperform the Dickey-Fuller-$\rho$ test). This corroborates the Chernoff-Savage property of the AHRT-$\phi$ test mentioned in Remark~\ref{remark:ChernofSavage}. 

\begin{figure}[!htb]
\centering
\hspace*{-33mm}
{\includegraphics[width=7.5in]{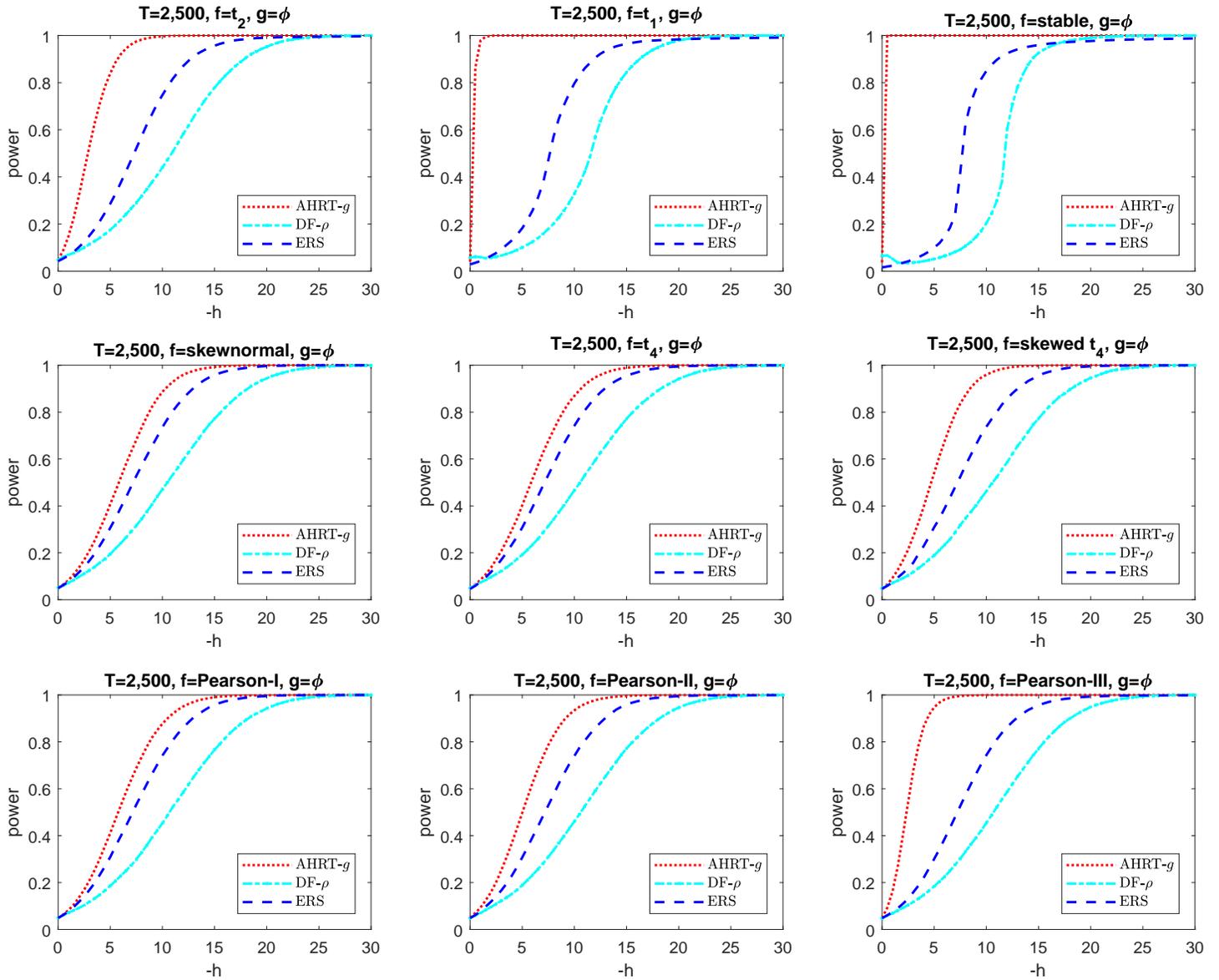}}
\caption{IID errors case: illustration of the Chernoff-Savage result. The figure shows large-sample power functions of the {\rm AHRT}-$\phi$ and other selected unit root tests under various true innovation densities $f$.}\label{figure_others_n2500}
\end{figure}

In order to investigate the Chernoff-Savage result for the AHRT-$\phi$ test even further, we consider in Figure~\ref{figure_others_n2500} the AHRT-$\phi$ test for i.i.d errors generated with nine different true innovation densities $f$. These include innovation densities $f$ that are extremely heavy-tailed, skewed, or both. The first row of graphs shows three extremely heavy-tailed distributions: Student $t_2$, Student $t_1$, and a stable distribution with parameter values $0.5$ for stability, $0$ for skewness, $1$ for scale, and $0$ location. As these densities do not all satisfy our maintained assumptions, these graphs exclude power envelopes and the AHRT-$\hat{f}$ power functions. The top three graphs in Figure~\ref{figure_others_n2500} show that the AHRT-$\phi$ is much more powerful than its competitors and that its power increases with the heaviness of the tail. The second and third row show the effect of skewness in $f$. Specifically, the AHRT-$\phi$'s power is higher when $f$ is skewed-normal (with skewness 0.8145) than that when $f$ is normal (in Figure~\ref{figure_n2500}). This indicates that the AHRT-$\phi$ can acquire power from skewness. The same conclusion can be drawn from the comparison of the AHRT-$\phi$ power function for $t_4$ and that of a skewed $t_4$ with skewness $\approx 2.7$. To further remove the effects of the other moments, in the third row, we also employ the Pearson distributions with identical mean, variance and kurtosis, but different skewness --- ${\rm skewness}=1$ for Pearson-I, ${\rm skewness}=3$ for Pearson-II, and ${\rm skewness}=6$ for Pearson-III. Comparing the corresponding power functions, it validates again that the larger the skewness of the true distribution $f$ is, the more powerful the ${\rm AHRT}^\phi$ becomes. 

A final remark on the size of the AHRT tests. In all cases where the true density $f$ satisfies our maintained assumption, i.e., $f\in\F$ (that is all cases in Figure~\ref{figure_n2500} and the skewnormal, $t_4$, Pearson-I, Pearson-II, and Pearson-III in Figure~\ref{figure_others_n2500}), the simulated sizes are between $4.9\%$ and $5.1\%$. This verifies the validity of the AHRTs claimed in Theorem~\ref{thm:propertiesAHRT}. In the other cases, i.e., $f\not\in\F$, the AHRT is somewhat conservative. More precisely, the simulated sizes of the AHRT-$\phi$ are $4.8\%$, $4.1\%$, $3.7\%$, and $4.7\%$ for the Student $t_2$, $t_1$, stable, and skew-$t_4$ distribution, respectively. This result seems consistent over all simulations. 

\subsubsection*{Small-sample performance} \label{sec:FinitePerform}

\begin{figure}[!htb]
\centering
\hspace*{-33mm}
{\includegraphics[width=7.5in]{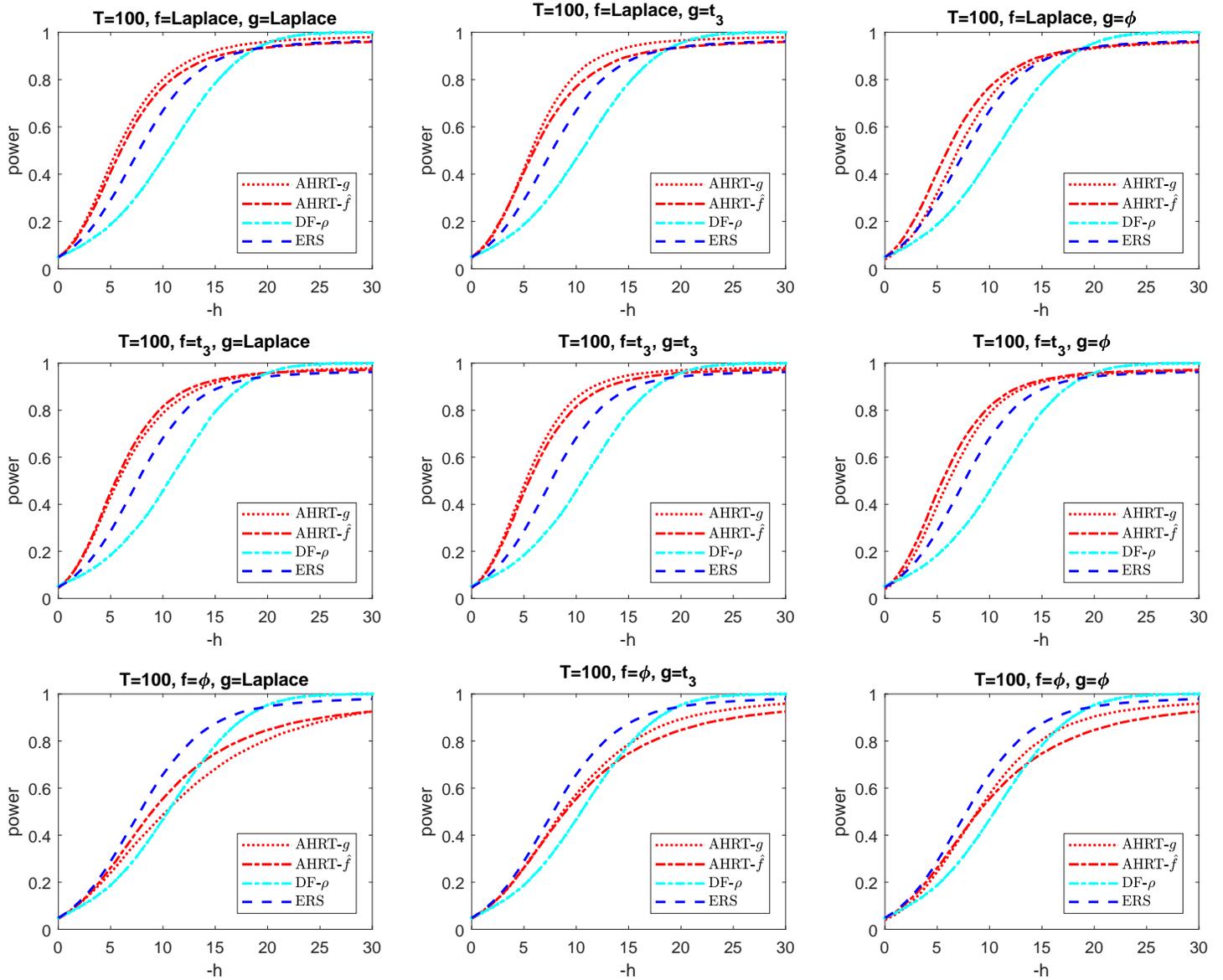}}
\caption{IID errors case: small-sample power functions of the {\rm AHRT}-g with various reference densities $g$, the {\rm AHRT}-$\hat{f}$, and other selected unit root tests under various true innovation densities: Gaussian, Laplace, Student $t_3$.}
\label{figure_n100}
\end{figure}

\begin{figure}[!htb]
\centering
\hspace*{-33mm}
{\includegraphics[width=7.5in]{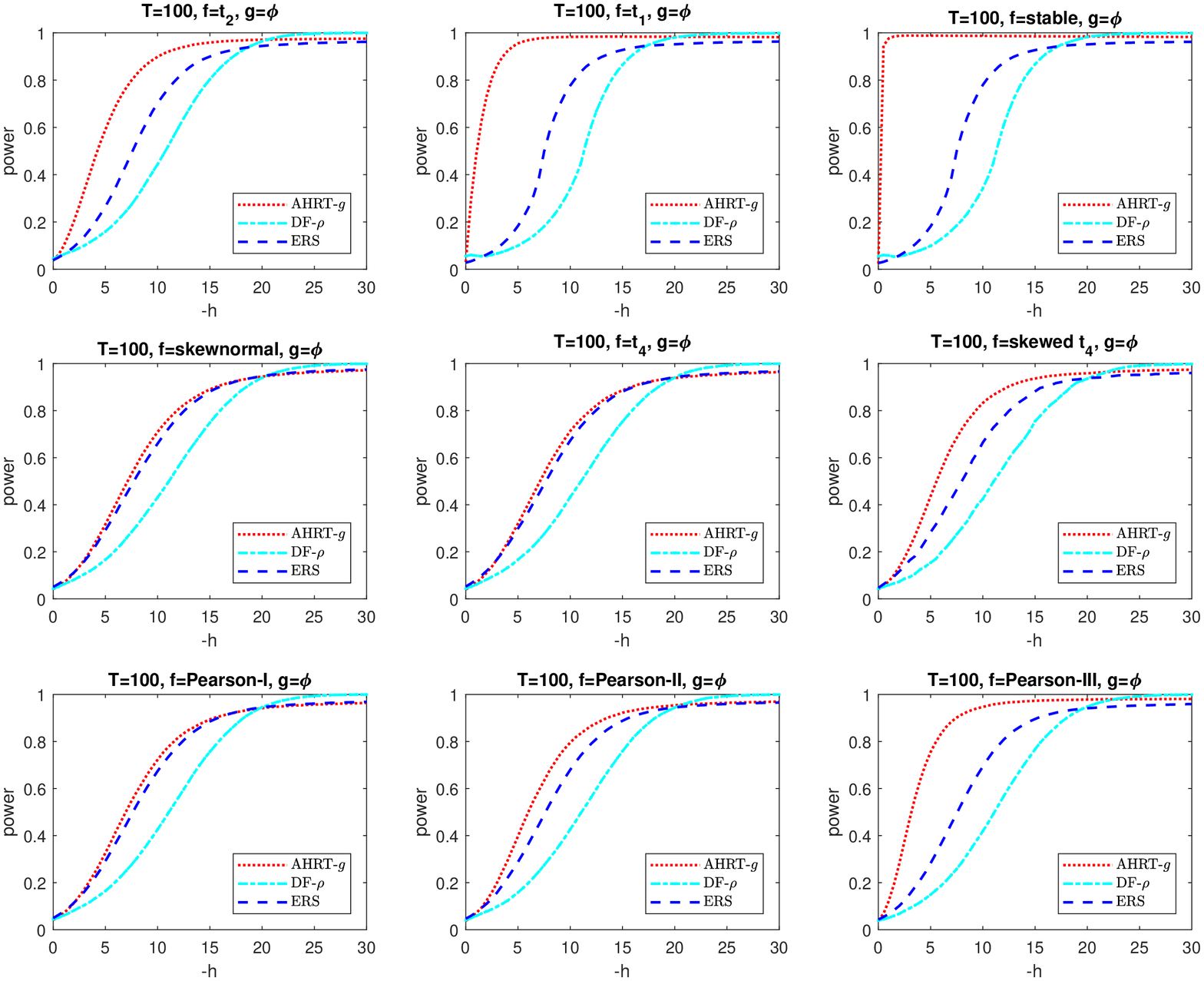}}
\caption{IID errors case: small-sample power functions of the Gaussian {\rm AHRT} and other selected unit root tests under some more true innovation densities.}
\label{figure_others_n100}
\end{figure}

We also report the performance of the AHRTs and the two competitors introduced above for smaller samples. Figures~\ref{figure_n100} and~\ref{figure_others_n100} are the small-sample versions, with $T = 100$, of Figures~\ref{figure_n2500} and~\ref{figure_others_n2500}, respectively. We observe that, even with a slight downward shift of the power functions for all three tests considered, the findings of the large-sample case remain valid in this small-sample case. For larger values of $h$, the DF-$\rho$ test sometimes dominates the other two tests. This is due to the fact that the DF-$\rho$ in this Monte-Carlo setting appears to have a superior convergence speed (towards its asymptotic power as sample size $T$ increases) to those of the AHRTs and the ERS test. This phenomenon appears only in (ultra)-small-sample cases, and disappears when $T$ gets larger (e.g., $T = 200$). In cases with enough samples (say, $T \geq 100$) and when $f$ is significantly away from the Gaussian density, irrespective of the choice of $g$, the AHRTs performs favorably. 

Concerning the small-sample size, we find it to range from about $4.0\%$ to $4.5\%$ for the cases where $f\in\F$. Again, when $f$ does not satisfy our maintained assumptions ($f\not\in\F$) the AHRT turns out to be conservative. More precisely, we find a size of $3.7\%$, $3.1\%$, $2.4\%$, and $4.1\%$ for the $t_2$, $t_1$, stable, and skew-$t_4$ distribution, respectively. This makes the improved power even more remarkable.

\begin{figure}[!htb]
\hspace*{-33mm}
\includegraphics[width=7.5in]{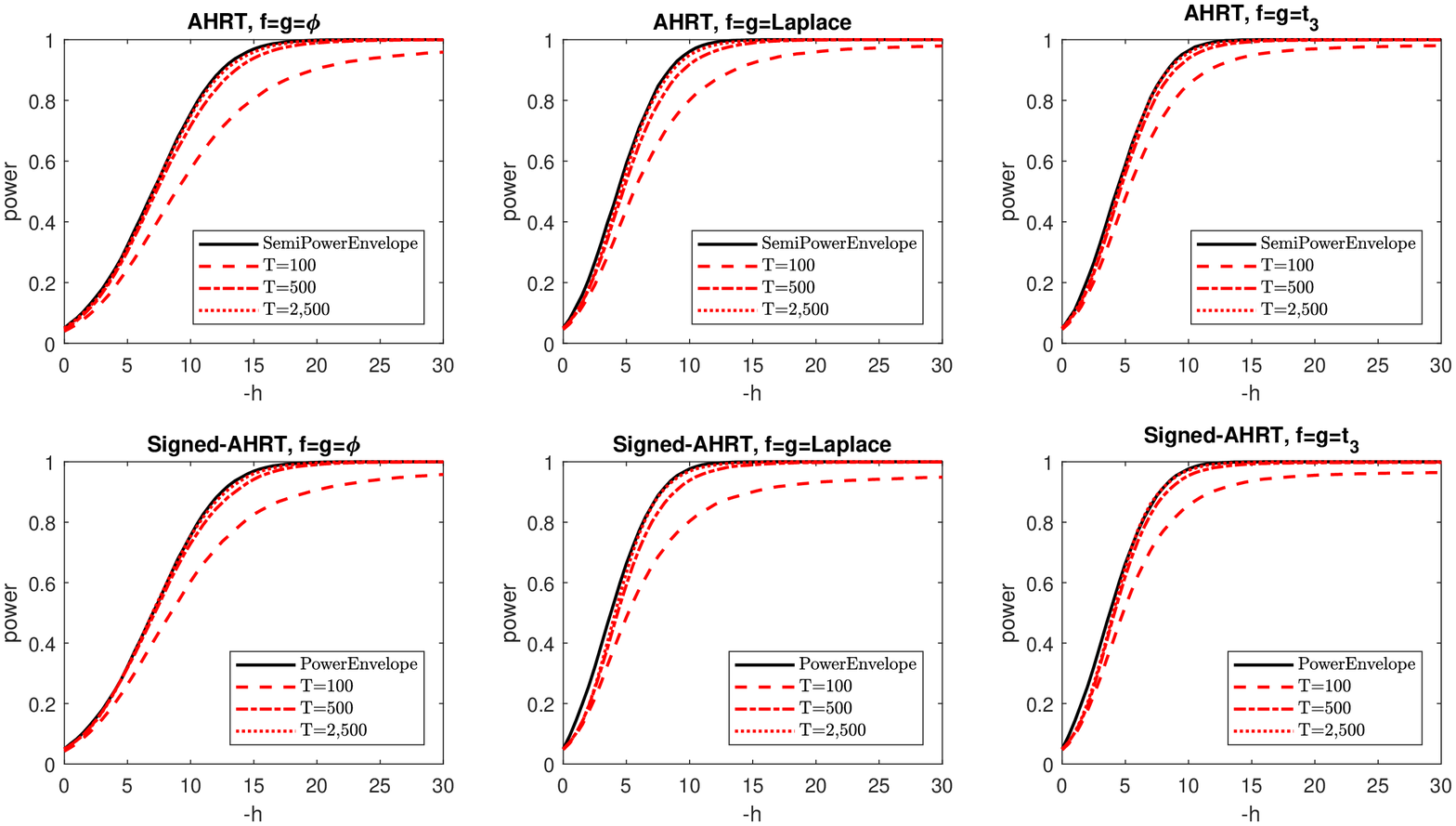}
\caption{IID errors case: power functions of {\rm AHRT}-$f$ ($g=f$) with different sample sizes.}
\label{figure_converge}
\end{figure}

It may also be useful to illustrate the convergence of the power function of the AHRT-${f}$ to the semiparametric power envelope as sample size $T$ increases. This is the purpose of Figure~\ref{figure_converge}. For three cases: Gaussian, Laplace, and Student $t_3$, we find that the convergence indeed occurs already at relatively small samples, which is not always the case for alternative unit root tests.

\subsubsection*{Performance of signed-rank-based AHRT test under restriction $f \in \Fs$}

In the case with the additional symmetric density restriction, $f \in \Fs$, we repeat the above Monte Carlo study above but for the signed-rank-based AHRT $\Ftf_g^\mathbb{S}(\bar{h},\alpha)$ introduced in Section~\ref{sec:approximatehybridranktests}. Not surprisingly, we find very similar results as above, except for a slightly higher power for all the AHRT tests (essentially due to the symmetry restriction imposed, in which case we have the adaptivity result). Since it's a bit unfair to compare with the competitor which actually does not benefit from this constraint, and also to conserve space, we put the associated simulation results --- two figures that can be treated as the signed-rank-based AHRT's counterparts of Figure~\ref{figure_n2500} and Figure~\ref{figure_n100} --- in the Appendix C in \ref{supplementA}. In short, the large-sample result shows that the power function of the signed AHRT-$f$ is tangent to the (parametric) power envelope, which corroborates the point-optimal property in Theorem~\ref{thm:propertiesAHRT} and in turn the adaptive result in Section~\ref{sec:invariance}; the Chernoff-Savage result still holds for the signed-rank version of the AHRT-$\phi$ test. The second figure therein shows that it works well in the small sample case. Furthermore, one can also check the convergence of the signed-rank-based AHRT-${f}$ power function to the parametric power envelope as sample size $T$ increases in Figure~\ref{figure_converge}.

\subsection{Simulation results with ARMA innocations} \label{sec:Montecarlo_ARMA}

\begin{figure}[!htb]
\hspace*{-33mm}
\includegraphics[width=7.5in]{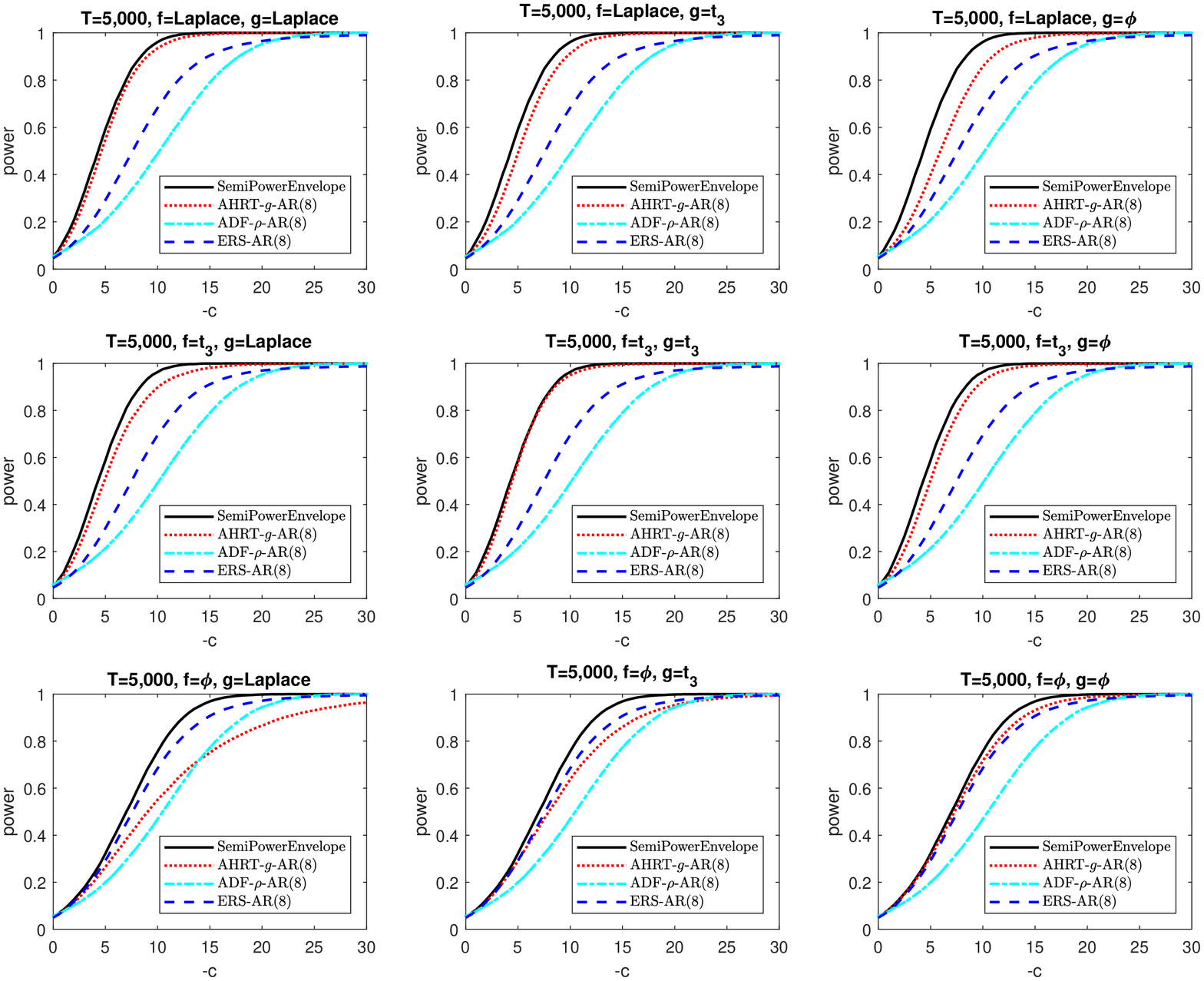}
\caption{ARMA errors case: large-sample power functions of the {\rm AHRT}-g with various reference densities $g$ and other selected unit root tests under the true innovation densities $f$: Gaussian, Laplace, Student's $t_3$.}
\label{figure_arma_n5000}
\end{figure}

\begin{figure}[!htb]
\hspace*{-33mm}
\includegraphics[width=7.5in]{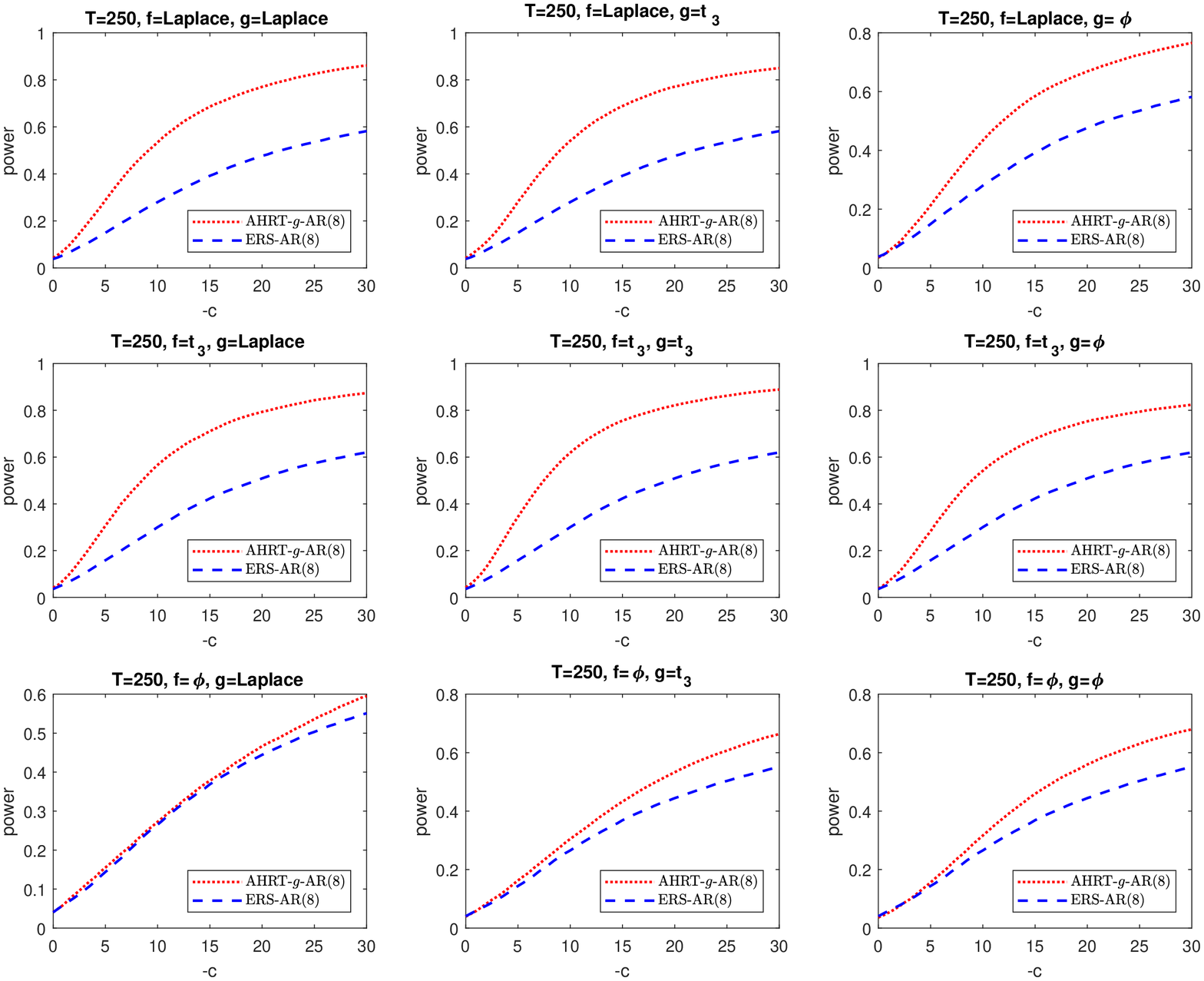}
\caption{ARMA errors case: small-sample power functions of the {\rm AHRT}-g with various reference densities $g$ and other selected unit root tests under various true innovation densities: Gaussian, Laplace, Student $t_3$.}
\label{figure_arma_n250}
\end{figure}

In this section we provide simulation results for the cases where the innovations follow an ARMA model, i.e., 
\begin{align*}
Y_t &= \mu + X_t,  \\
X_t &= \rho X_{t-1}+v_t, \\
v_t &= -0.5 v_{t-1} + \varepsilon_t - 0.5 \varepsilon_{t-1}, \quad t\in\SN,
\end{align*}
where the assumption on $\varepsilon_t$ stay unchanged. This corresponds to a lag-polynomial $\Gamma(L)$ with an infinite order, which we employ here in order to show by simulations the potential that the finite order $p$ restriction could be relaxed. As for the estimation of $\widehat{\Gamma}(L)$ in (\ref{eqn:arma_regression}) used by the AHRT test and its competitors, we fix the AR regression order $p = 8$. Since there exists efficiency loss in the estimation of error autocorrelation parameters $\Gamma_1,\dots,\Gamma_p$ (for finite-sample cases), here we increase the sample sizes for both the large-sample case (to $T = 5,000$) and the small-sample case (to $T = 250$). Under the same organization of true and reference densities as for the i.i.d.\ case, we provide the simulation results for large-sample case and small-sample case in Figure~\ref{figure_arma_n5000} and Figure~\ref{figure_arma_n250}, respectively. Note that, in Figure~\ref{figure_arma_n250}, we remove the powers of the ADF-$\rho$ test since it is severely oversized.

From these results, we draw similar conclusions to those in the i.i.d.\ case, except for a slight power loss for all these chosen unit root tests. These results validate that, on one hand, the serial correlation in the errors can be well handled, as for the cases of other unit root tests, with an additional auto-regression for the increments of the observed process. On the other hand, they also indicate that even though in the limit the unknown autocorrelation structure has no effects on the inference for $\rho$, its estimation consumes some efficiency in finite sample cases. Yet, Figure~\ref{figure_arma_n250} shows that the AHRT suffers less from this kind of efficiency loss or, in other words, the AHRT has a faster convergence speed of its power to the associated envelope as $T\to\infty$. 

\section{Conclusion}\label{sec:conclusion}
\noindent
This paper has provided a structural representation of the limit experiment of the standard unit root model in a univariate but semiparametric setting. Using invariance arguments, we have derived the semiparametric power envelope. These invariance structures also lead, using the Neyman-Pearson lemma, to point-optimal semiparametric tests. The analysis naturally leads to the use of rank-based statistics.

Our tests are asymptotically valid, invariant, and (with a correctly chosen reference density) point-optimal. Moreover, we establish a Chernoff-Savage type property of our test: irrespective of the reference density chosen, our test outperforms its classical competitor which in this case is the ERS test. Finally, we introduced a simplified version of our test and show, in a Monte-Carlo study, that our theoretical results carry over to finite samples.

As potential future work we mention the use of similar ideas to construct hybrid rank-based tests in more general time-series models with, for instance, a deterministic time trend term, or stochastic volatility. Also, the structural representation of the limit experiment and its invariance properties could be applied to other non-stationary time-series models, for instance, cointegration and predictive regression models.

\begin{supplement}
\sname{Supplement A}\label{supplementA}
\stitle{Supplement to ``Semiparametrically optimal hybrid rank tests for unit roots''}
\sdescription{This supplemental file contains technical proofs for propositions and theorems in the main context.}
\end{supplement}

\bibliographystyle{imsart-nameyear}
\bibliography{references}

\begin{thebibliography}{39}

\bibitem[\protect\citeauthoryear{Ahn, Fotopoulos and
  He}{2001}]{AhnFotopoulosHe03}
\begin{barticle}[author]
\bauthor{\bsnm{Ahn},~\bfnm{Sung~K}\binits{S.~K.}},
  \bauthor{\bsnm{Fotopoulos},~\bfnm{Stergios~B}\binits{S.~B.}} \AND
  \bauthor{\bsnm{He},~\bfnm{Lijian}\binits{L.}}
(\byear{2001}).
\btitle{Unit root tests with infinite variance errors}.
\bjournal{Econometric Reviews}
\bvolume{20}
\bpages{461--483}.
\end{barticle}
\endbibitem

\bibitem[\protect\citeauthoryear{Bickel}{1982}]{Bickel1982}
\begin{barticle}[author]
\bauthor{\bsnm{Bickel},~\bfnm{Peter~J}\binits{P.~J.}}
(\byear{1982}).
\btitle{On adaptive estimation}.
\bjournal{The Annals of Statistics}
\bpages{647--671}.
\end{barticle}
\endbibitem

\bibitem[\protect\citeauthoryear{Bickel et~al.}{1998}]{BKRW}
\begin{bbook}[author]
\bauthor{\bsnm{Bickel},~\bfnm{Peter~J}\binits{P.~J.}},
  \bauthor{\bsnm{Klaassen},~\bfnm{Chris~A}\binits{C.~A.}},
  \bauthor{\bsnm{Bickel},~\bfnm{Peter~J}\binits{P.~J.}},
  \bauthor{\bsnm{Ritov},~\bfnm{Y}\binits{Y.}},
  \bauthor{\bsnm{Klaassen},~\bfnm{J}\binits{J.}},
  \bauthor{\bsnm{Wellner},~\bfnm{Jon~A}\binits{J.~A.}} \AND
  \bauthor{\bsnm{Ritov},~\bfnm{YA'Acov}\binits{Y.}}
(\byear{1998}).
\btitle{Efficient and adaptive estimation for semiparametric models}
\bvolume{2}.
\bpublisher{Springer New York}.
\end{bbook}
\endbibitem

\bibitem[\protect\citeauthoryear{Brockwell and
  Davis}{2016}]{BrockwellDavis2002}
\begin{bbook}[author]
\bauthor{\bsnm{Brockwell},~\bfnm{Peter~J}\binits{P.~J.}} \AND
  \bauthor{\bsnm{Davis},~\bfnm{Richard~A}\binits{R.~A.}}
(\byear{2016}).
\btitle{Introduction to time series and forecasting}.
\bpublisher{springer}.
\end{bbook}
\endbibitem

\bibitem[\protect\citeauthoryear{Callegari, Cappuccio and
  Lubian}{2003}]{CallegariCappuccioLubian03}
\begin{barticle}[author]
\bauthor{\bsnm{Callegari},~\bfnm{Francesca}\binits{F.}},
  \bauthor{\bsnm{Cappuccio},~\bfnm{Nunzio}\binits{N.}} \AND
  \bauthor{\bsnm{Lubian},~\bfnm{Diego}\binits{D.}}
(\byear{2003}).
\btitle{Asymptotic inference in time series regressions with a unit root and
  infinite variance errors}.
\bjournal{Journal of Statistical Planning and Inference}
\bvolume{116}
\bpages{277--303}.
\end{barticle}
\endbibitem

\bibitem[\protect\citeauthoryear{Cassart, Hallin and
  Paindaveine}{2010}]{CassartHallinPain2010}
\begin{bincollection}[author]
\bauthor{\bsnm{Cassart},~\bfnm{Delphine}\binits{D.}},
  \bauthor{\bsnm{Hallin},~\bfnm{Marc}\binits{M.}} \AND
  \bauthor{\bsnm{Paindaveine},~\bfnm{Davy}\binits{D.}}
(\byear{2010}).
\btitle{On the estimation of cross-information quantities in rank-based
  inference}.
In \bbooktitle{Nonparametrics and Robustness in Modern Statistical Inference
  and Time Series Analysis: A Festschrift in Honor of Professor Jana
  Jure{\v{c}}kov{\'a}}
\bpages{35--45}.
\bpublisher{Institute of Mathematical Statistics}.
\end{bincollection}
\endbibitem

\bibitem[\protect\citeauthoryear{Chan and Wei}{1988}]{ChanWei88}
\begin{barticle}[author]
\bauthor{\bsnm{Chan},~\bfnm{Ngai~Hang}\binits{N.~H.}} \AND
  \bauthor{\bsnm{Wei},~\bfnm{CZ}\binits{C.}}
(\byear{1988}).
\btitle{Limiting distributions of least squares estimates of unstable
  autoregressive processes}.
\bjournal{The Annals of Statistics}
\bpages{367--401}.
\end{barticle}
\endbibitem

\bibitem[\protect\citeauthoryear{Chernoff and
  Savage}{1958}]{ChernoffSavage1958}
\begin{barticle}[author]
\bauthor{\bsnm{Chernoff},~\bfnm{Herman}\binits{H.}} \AND
  \bauthor{\bsnm{Savage},~\bfnm{I~Richard}\binits{I.~R.}}
(\byear{1958}).
\btitle{Asymptotic normality and efficiency of certain nonparametric test
  statistics}.
\bjournal{The Annals of Mathematical Statistics}
\bpages{972--994}.
\end{barticle}
\endbibitem

\bibitem[\protect\citeauthoryear{Choi}{2015}]{Choi2015}
\begin{barticle}[author]
\bauthor{\bsnm{Choi},~\bfnm{In}\binits{I.}}
(\byear{2015}).
\btitle{Almost all about unit roots}.
\bjournal{Cambridge Books}.
\end{barticle}
\endbibitem

\bibitem[\protect\citeauthoryear{Dickey and Fuller}{1979}]{DickeyFuller79}
\begin{barticle}[author]
\bauthor{\bsnm{Dickey},~\bfnm{David~A}\binits{D.~A.}} \AND
  \bauthor{\bsnm{Fuller},~\bfnm{Wayne~A}\binits{W.~A.}}
(\byear{1979}).
\btitle{Distribution of the estimators for autoregressive time series with a
  unit root}.
\bjournal{Journal of the American statistical association}
\bvolume{74}
\bpages{427--431}.
\end{barticle}
\endbibitem

\bibitem[\protect\citeauthoryear{Dickey and Fuller}{1981}]{DickeyFuller81}
\begin{barticle}[author]
\bauthor{\bsnm{Dickey},~\bfnm{DA}\binits{D.}} \AND
  \bauthor{\bsnm{Fuller},~\bfnm{WA}\binits{W.}}
(\byear{1981}).
\btitle{Likelihood ratio statistics for autoregressive time series with a unit
  root}.
\bjournal{Econometrica}
\bvolume{49}
\bpages{1057--1072}.
\end{barticle}
\endbibitem

\bibitem[\protect\citeauthoryear{Dufour and King}{1991}]{DufourKing91}
\begin{barticle}[author]
\bauthor{\bsnm{Dufour},~\bfnm{Jean-Marie}\binits{J.-M.}} \AND
  \bauthor{\bsnm{King},~\bfnm{Maxwell~L}\binits{M.~L.}}
(\byear{1991}).
\btitle{Optimal invariant tests for the autocorrelation coefficient in linear
  regressions with stationary or nonstationary AR (1) errors}.
\bjournal{Journal of Econometrics}
\bvolume{47}
\bpages{115--143}.
\end{barticle}
\endbibitem

\bibitem[\protect\citeauthoryear{Elliott and
  M{\"u}ller}{2006}]{ElliottMuller2006}
\begin{barticle}[author]
\bauthor{\bsnm{Elliott},~\bfnm{Graham}\binits{G.}} \AND
  \bauthor{\bsnm{M{\"u}ller},~\bfnm{Ulrich~K}\binits{U.~K.}}
(\byear{2006}).
\btitle{Minimizing the impact of the initial condition on testing for unit
  roots}.
\bjournal{Journal of Econometrics}
\bvolume{135}
\bpages{285--310}.
\end{barticle}
\endbibitem

\bibitem[\protect\citeauthoryear{Elliott, Rothenberg and Stock}{1996}]{ERS1996}
\begin{barticle}[author]
\bauthor{\bsnm{Elliott},~\bfnm{Graham}\binits{G.}},
  \bauthor{\bsnm{Rothenberg},~\bfnm{Thomas~J}\binits{T.~J.}} \AND
  \bauthor{\bsnm{Stock},~\bfnm{James~H}\binits{J.~H.}}
(\byear{1996}).
\btitle{Efficient Tests for an Autoregressive Unit Root}.
\bjournal{Econometrica}
\bvolume{64}
\bpages{813--836}.
\end{barticle}
\endbibitem

\bibitem[\protect\citeauthoryear{H{\'a}jek and
  Sid{\'a}k}{1967}]{HajekSidak1967}
\begin{barticle}[author]
\bauthor{\bsnm{H{\'a}jek},~\bfnm{Jaroslav}\binits{J.}} \AND
  \bauthor{\bsnm{Sid{\'a}k},~\bfnm{Zbynek}\binits{Z.}}
(\byear{1967}).
\btitle{Theory of rank tests.}
\end{barticle}
\endbibitem

\bibitem[\protect\citeauthoryear{Hallin and Puri}{1988}]{HallinPuri1988}
\begin{barticle}[author]
\bauthor{\bsnm{Hallin},~\bfnm{Marc}\binits{M.}} \AND
  \bauthor{\bsnm{Puri},~\bfnm{Madan~L}\binits{M.~L.}}
(\byear{1988}).
\btitle{Optimal rank-based procedures for time series analysis: testing an ARMA
  model against other ARMA models}.
\bjournal{The Annals of Statistics}
\bpages{402--432}.
\end{barticle}
\endbibitem

\bibitem[\protect\citeauthoryear{Hallin and Puri}{1994}]{HallinPuri1994}
\begin{barticle}[author]
\bauthor{\bsnm{Hallin},~\bfnm{Marc}\binits{M.}} \AND
  \bauthor{\bsnm{Puri},~\bfnm{Madan~L}\binits{M.~L.}}
(\byear{1994}).
\btitle{Aligned rank tests for linear models with autocorrelated error terms}.
\bjournal{Journal of Multivariate Analysis}
\bvolume{50}
\bpages{175--237}.
\end{barticle}
\endbibitem

\bibitem[\protect\citeauthoryear{Hallin, Van~den Akker and
  Werker}{2011}]{HvdAW1}
\begin{barticle}[author]
\bauthor{\bsnm{Hallin},~\bfnm{Marc}\binits{M.}}, \bauthor{\bparticle{Van~den}
  \bsnm{Akker},~\bfnm{Ramon}\binits{R.}} \AND \bauthor{\bsnm{Werker},~\bfnm{Bas
  J~M}\binits{B.~J.~M.}}
(\byear{2011}).
\btitle{A class of simple distribution-free rank-based unit root tests}.
\bjournal{Journal of econometrics}
\bvolume{163}
\bpages{200--214}.
\end{barticle}
\endbibitem

\bibitem[\protect\citeauthoryear{Hallin, Van Den~Akker and
  Werker}{2015}]{HvdAW2}
\begin{bincollection}[author]
\bauthor{\bsnm{Hallin},~\bfnm{Marc}\binits{M.}}, \bauthor{\bsnm{Van
  Den~Akker},~\bfnm{Ramon}\binits{R.}} \AND \bauthor{\bsnm{Werker},~\bfnm{Bas
  J~M}\binits{B.~J.~M.}}
(\byear{2015}).
\btitle{On quadratic expansions of log-likelihoods and a general asymptotic
  linearity result}.
In \bbooktitle{Mathematical Statistics and Limit Theorems}
\bpages{147--165}.
\bpublisher{Springer}.
\end{bincollection}
\endbibitem

\bibitem[\protect\citeauthoryear{Hasan}{2001}]{Hasan2001}
\begin{barticle}[author]
\bauthor{\bsnm{Hasan},~\bfnm{Mohammad~N}\binits{M.~N.}}
(\byear{2001}).
\btitle{Rank tests of unit root hypothesis with infinite variance errors}.
\bjournal{Journal of Econometrics}
\bvolume{104}
\bpages{49--65}.
\end{barticle}
\endbibitem

\bibitem[\protect\citeauthoryear{Jansson}{2008}]{Jansson2008}
\begin{barticle}[author]
\bauthor{\bsnm{Jansson},~\bfnm{Michael}\binits{M.}}
(\byear{2008}).
\btitle{Semiparametric power envelopes for tests of the unit root hypothesis}.
\bjournal{Econometrica}
\bvolume{76}
\bpages{1103--1142}.
\end{barticle}
\endbibitem

\bibitem[\protect\citeauthoryear{Jeganathan}{1995}]{Jeganathan1995}
\begin{barticle}[author]
\bauthor{\bsnm{Jeganathan},~\bfnm{Pradeep}\binits{P.}}
(\byear{1995}).
\btitle{Some aspects of asymptotic theory with applications to time series
  models}.
\bjournal{Econometric Theory}
\bvolume{11}
\bpages{818--887}.
\end{barticle}
\endbibitem

\bibitem[\protect\citeauthoryear{Jeganathan}{1997}]{Jeganathan1997}
\begin{barticle}[author]
\bauthor{\bsnm{Jeganathan},~\bfnm{P}\binits{P.}}
(\byear{1997}).
\btitle{On asymptotic inference in linear cointegrated time series systems}.
\bjournal{Econometric Theory}
\bvolume{13}
\bpages{692--745}.
\end{barticle}
\endbibitem

\bibitem[\protect\citeauthoryear{Klaassen}{1987}]{Klaassen1987}
\begin{barticle}[author]
\bauthor{\bsnm{Klaassen},~\bfnm{Chris~AJ}\binits{C.~A.}}
(\byear{1987}).
\btitle{Consistent estimation of the influence function of locally
  asymptotically linear estimators}.
\bjournal{The Annals of Statistics}
\bpages{1548--1562}.
\end{barticle}
\endbibitem

\bibitem[\protect\citeauthoryear{Kreiss}{1987}]{Kreiss1987}
\begin{barticle}[author]
\bauthor{\bsnm{Kreiss},~\bfnm{Jens-Peter}\binits{J.-P.}}
(\byear{1987}).
\btitle{On adaptive estimation in stationary ARMA processes}.
\bjournal{The Annals of Statistics}
\bpages{112--133}.
\end{barticle}
\endbibitem

\bibitem[\protect\citeauthoryear{Le~Cam}{2012}]{LeCam1986}
\begin{bbook}[author]
\bauthor{\bsnm{Le~Cam},~\bfnm{Lucien}\binits{L.}}
(\byear{2012}).
\btitle{Asymptotic methods in statistical decision theory}.
\bpublisher{Springer Science \& Business Media}.
\end{bbook}
\endbibitem

\bibitem[\protect\citeauthoryear{M{\"u}ller}{2011}]{Muller2011}
\begin{barticle}[author]
\bauthor{\bsnm{M{\"u}ller},~\bfnm{Ulrich~K}\binits{U.~K.}}
(\byear{2011}).
\btitle{Efficient tests under a weak convergence assumption}.
\bjournal{Econometrica}
\bvolume{79}
\bpages{395--435}.
\end{barticle}
\endbibitem

\bibitem[\protect\citeauthoryear{M{\"u}ller and
  Elliott}{2003}]{MullerElliott2003}
\begin{barticle}[author]
\bauthor{\bsnm{M{\"u}ller},~\bfnm{Ulrich~K}\binits{U.~K.}} \AND
  \bauthor{\bsnm{Elliott},~\bfnm{Graham}\binits{G.}}
(\byear{2003}).
\btitle{Tests for unit roots and the initial condition}.
\bjournal{Econometrica}
\bvolume{71}
\bpages{1269--1286}.
\end{barticle}
\endbibitem

\bibitem[\protect\citeauthoryear{M{\"u}ller and
  Watson}{2008}]{MullerWatson2008}
\begin{barticle}[author]
\bauthor{\bsnm{M{\"u}ller},~\bfnm{Ulrich~K}\binits{U.~K.}} \AND
  \bauthor{\bsnm{Watson},~\bfnm{Mark~W}\binits{M.~W.}}
(\byear{2008}).
\btitle{Testing Models of Low-Frequency Variability}.
\bjournal{Econometrica}
\bvolume{76}
\bpages{979--1016}.
\end{barticle}
\endbibitem

\bibitem[\protect\citeauthoryear{Patterson}{2011}]{Patterson2011}
\begin{bbook}[author]
\bauthor{\bsnm{Patterson},~\bfnm{Kerry}\binits{K.}}
(\byear{2011}).
\btitle{Unit root tests in time series volume 1: key concepts and problems}
\bvolume{1}.
\bpublisher{Palgrave Macmillan}.
\end{bbook}
\endbibitem

\bibitem[\protect\citeauthoryear{Patterson}{2012}]{Patterson2012}
\begin{bbook}[author]
\bauthor{\bsnm{Patterson},~\bfnm{Kerry}\binits{K.}}
(\byear{2012}).
\btitle{Unit root tests in time series volume 2: extensions and developments}
\bvolume{2}.
\bpublisher{Palgrave Macmillan}.
\end{bbook}
\endbibitem

\bibitem[\protect\citeauthoryear{Phillips}{1987}]{Phillips1987}
\begin{barticle}[author]
\bauthor{\bsnm{Phillips},~\bfnm{Peter~CB}\binits{P.~C.}}
(\byear{1987}).
\btitle{Time series regression with a unit root}.
\bjournal{Econometrica}
\bpages{277--301}.
\end{barticle}
\endbibitem

\bibitem[\protect\citeauthoryear{Phillips and Perron}{1988}]{PhillipsPerron88}
\begin{barticle}[author]
\bauthor{\bsnm{Phillips},~\bfnm{Peter~CB}\binits{P.~C.}} \AND
  \bauthor{\bsnm{Perron},~\bfnm{Pierre}\binits{P.}}
(\byear{1988}).
\btitle{Testing for a unit root in time series regression}.
\bjournal{Biometrika}
\bvolume{75}
\bpages{335--346}.
\end{barticle}
\endbibitem

\bibitem[\protect\citeauthoryear{Rothenberg and
  Stock}{1997}]{RothenbergStock97}
\begin{barticle}[author]
\bauthor{\bsnm{Rothenberg},~\bfnm{Thomas~J}\binits{T.~J.}} \AND
  \bauthor{\bsnm{Stock},~\bfnm{James~H}\binits{J.~H.}}
(\byear{1997}).
\btitle{Inference in a nearly integrated autoregressive model with nonnormal
  innovations}.
\bjournal{Journal of Econometrics}
\bvolume{80}
\bpages{269--286}.
\end{barticle}
\endbibitem

\bibitem[\protect\citeauthoryear{Rudin}{1987}]{Rudin1987}
\begin{barticle}[author]
\bauthor{\bsnm{Rudin},~\bfnm{Walter}\binits{W.}}
(\byear{1987}).
\btitle{Real and complex analysis}.
\end{barticle}
\endbibitem

\bibitem[\protect\citeauthoryear{Saikkonen and
  Luukkonen}{1993}]{SaikkonenLuukkonen1993}
\begin{barticle}[author]
\bauthor{\bsnm{Saikkonen},~\bfnm{Pentti}\binits{P.}} \AND
  \bauthor{\bsnm{Luukkonen},~\bfnm{Ritva}\binits{R.}}
(\byear{1993}).
\btitle{Point optimal tests for testing the order of differencing in ARIMA
  models}.
\bjournal{Econometric Theory}
\bvolume{9}
\bpages{343--362}.
\end{barticle}
\endbibitem

\bibitem[\protect\citeauthoryear{Schick}{1986}]{Schick1986}
\begin{barticle}[author]
\bauthor{\bsnm{Schick},~\bfnm{Anton}\binits{A.}}
(\byear{1986}).
\btitle{On asymptotically efficient estimation in semiparametric models}.
\bjournal{The Annals of Statistics}
\bpages{1139--1151}.
\end{barticle}
\endbibitem

\bibitem[\protect\citeauthoryear{van~der Vaart}{2000}]{vdVaart00}
\begin{bbook}[author]
\bauthor{\bparticle{van~der} \bsnm{Vaart},~\bfnm{AW}\binits{A.}}
(\byear{2000}).
\btitle{Asymptotic Statistics}
\bvolume{3}.
\bpublisher{Cambridge University Press}.
\end{bbook}
\endbibitem

\bibitem[\protect\citeauthoryear{White}{1958}]{White58}
\begin{barticle}[author]
\bauthor{\bsnm{White},~\bfnm{John~S}\binits{J.~S.}}
(\byear{1958}).
\btitle{The limiting distribution of the serial correlation coefficient in the
  explosive case}.
\bjournal{The Annals of Mathematical Statistics}
\bpages{1188--1197}.
\end{barticle}
\endbibitem

\end{thebibliography}

\end{document}